\documentclass{sig-alternate-05-2015}

\usepackage[T1]{fontenc}

\usepackage[
  pdftitle={POPE: Partial Order Preserving Encoding},
  pdfauthor={Daniel S. Roche, Daniel Apon, Seung Geol Choi, Arkady
    Yerukhimovich},
  hidelinks,
  bookmarksnumbered=true,
  pdftex,
]{hyperref}
\usepackage[usenames,dvipsnames]{xcolor}
\newtheorem{definition}{Definition}
\newtheorem{theorem}{Theorem}

\usepackage{times}
\usepackage[lined,boxruled,commentsnumbered]{algorithm2e}
\usepackage{framed}
\usepackage[capitalise,noabbrev]{cleveref}

\usepackage{xspace}
\usepackage{url}
\usepackage{graphicx}
\usepackage{fancyvrb}
\usepackage{multirow,multicol}
\usepackage{enumitem}
\setlist[itemize]{itemsep=0pt}
\setlist[enumerate]{itemsep=0pt}

\usepackage[font=small,labelfont=bf]{caption}

\usepackage{xcolor}

\newcommand{\BT}{\begin{theorem}}
\newcommand{\ET}{\end{theorem}}

\newcommand{\BD}{\begin{definition}}
\newcommand{\ED}{\end{definition}}

\newcommand{\BCR}{\begin{corollary}}
\newcommand{\ECR}{\end{corollary}}

\newcommand{\BEX}{\begin{example}}
\newcommand{\EEX}{\end{example}}

\newcommand{\BL}{\begin{lemma}}
\newcommand{\EL}{\end{lemma}}

\newcommand{\BP}{\begin{proposition}}
\newcommand{\EP}{\end{proposition}}

\newcommand{\BCM}{\begin{claim}}
\newcommand{\ECM}{\end{claim}}

\newcommand{\BPF}{\begin{proof}}
\newcommand{\EPF}{\end{proof}}
\newcommand{\BEN}{\begin{enumerate}}
\newcommand{\EEN}{\end{enumerate}}

\newcommand{\BI}{\begin{itemize}}
\newcommand{\EI}{\end{itemize}}

\newcommand{\BO}{\begin{observation}}
\newcommand{\EO}{\end{observation}}

\newcommand{\BDS}{\begin{description}}
\newcommand{\EDS}{\end{description}}

\def\AAA{{\mathcal A}}     %
\def\BBB{{\mathcal B}}     %
\def\TTT{{\mathcal T}}     %
\def\zo{{\{0,1\}}}    %
\def\to{{\rightarrow}}
\def\from{{\,\leftarrow\,}}
\def\xor{{\oplus}}         %

\def\view{{\textsc{view}}}

\renewcommand{\>}{{\rangle}}

\def\squareforqed{\hbox{\(\blacksquare\)}}
\def\qed{\hfill \squareforqed}

\def\HYBRID{\textsc{hybrid}}

\newcommand{\Sim}{\ensuremath{\mathsf{Sim}}}

\newcommand{\secparam}{\ensuremath{\lambda}}

\newcommand{\ignore}[1]{}
\def\A{\mathcal{A}}

\def\Enc{\mathsf{Enc}}

\def\VEnc{\mathsf{vEnc}}

\def\gen{{\sf gen}}
\def\enc{{\sf enc}}
\def\dec{{\sf dec}}

\newcommand{\negl}{{\sf negl}}

\def\S{\ensuremath{\mathsf{S}}}

\newcounter{defcounter}
\newlength{\protowidth}

\newcommand{\advan}{\ensuremath{\mathbf{Adv}}}

\newcommand{\olrk}[1]{%
   \ifx\nursymbol#1\else\!\!\mskip4.5mu plus 0.5mu\left(#1\right)\fi}
\newcommand{\elrk}[1]{%
   \ifx\nursymbol#1\else%
        \!\!\mskip4.5mu plus0.5mu\left[\mskip2.5mu plus0.5mu #1\right]\fi}

\def\S{\ensuremath{\mathsf{S}}}

\def\rng{\ensuremath{{\mathsf{ind\mbox{-}ocpa}}}\xspace}

\def\POPE{\mathsf{POPE}}

\def\exprngz{\mathsf{EXP}_{\AAA}^{\rng}(\POPE, \secparam, 0)}
\def\exprngo{\mathsf{EXP}_{\AAA}^{\rng}(\POPE, \secparam, 1)}

\def\dcpa{\mathsf{ind\mbox{-}dcpa}}
\def\expdcpa{\mathsf{EXP}_{\AAA}^{\dcpa}(\Pi, \secparam, b)}
\def\expdcpaz{\mathsf{EXP}_{\AAA}^{\dcpa}(\Pi, \secparam, 0)}
\def\expdcpao{\mathsf{EXP}_{\AAA}^{\dcpa}(\Pi, \secparam, 1)}
\def\advdcpa{\advan_\AAA^\dcpa(\Pi,\secparam)}

\renewcommand{\paragraph}[1]{\smallskip \noindent \textbf{#1}~}
\usepackage{etoolbox}

\def\path{\ensuremath{{\sf location}}}

\def\ins{\mathsf{insert}}
\def\range{\mathsf{range}}
\def\profile{\mathsf{profile}}
\def\seq{\mathsf{seq}}
\def\labels{\mathsf{labels}}
\def\info{\mathsf{info}}

\newcommand{\cl}{{\color{Blue} \ensuremath{\mathsf{Cl}}}\xspace}
\newcommand{\ser}{{\color{BrickRed} \ensuremath{\mathsf{Ser}}}\xspace}
\newcommand{\Psetup}{\ensuremath{\mathsf{Setup}}\xspace}
\newcommand{\Pinsert}{\ensuremath{\mathsf{Insert}}\xspace}
\newcommand{\Psearch}{\ensuremath{\mathsf{Search}}\xspace}
\newcommand{\Psplit}{\ensuremath{\mathsf{Split}}\xspace}
\newcommand{\ellL}{\ensuremath{\ell_{\mathsf{left}}}\xspace}
\newcommand{\ellR}{\ensuremath{\ell_{\mathsf{right}}}\xspace}

\begin{document}

\CopyrightYear{2016}
\setcopyright{licensedusgovmixed}
\conferenceinfo{CCS'16,}{October 24 - 28, 2016, Vienna, Austria}
\isbn{978-1-4503-4139-4/16/10}\acmPrice{\$15.00}
\doi{http://dx.doi.org/10.1145/2976749.2978345}

\title{POPE: Partial Order Preserving Encoding}

\numberofauthors{1}

\author{
\alignauthor
Daniel S. Roche*,\quad
Daniel Apon$^\dagger$,\quad
Seung Geol Choi*,\quad
Arkady Yerukhimovich$^\ddagger$ 
\and
\affaddr{*United States Naval Academy, Annapolis, Maryland, USA} 
\\
\affaddr{$^\dagger$University of Maryland, College Park, Maryland, USA}
\\
\affaddr{$^\ddagger$MIT Lincoln Laboratory, Lexington, Massachusetts, USA}
\and
\email{
  \{roche,choi\}@usna.edu,
  dapon@cs.umd.edu,
  arkady@ll.mit.edu
}
}

\maketitle 

\begin{abstract}

Recently there has been much interest in performing search queries over
encrypted data to enable functionality while protecting sensitive data. 
One particularly efficient mechanism for executing such queries is
order-preserving encryption/encoding (OPE) which results in ciphertexts that
preserve the relative order of the underlying plaintexts thus allowing range and
comparison queries to be performed directly on ciphertexts.  
Recently, Popa et al.~(S\&P 2013) gave the first construction of an
ideally-secure OPE scheme and Kerschbaum~(CCS 2015) showed how to
achieve the even
stronger notion of frequency-hiding OPE.  However, as 
Naveed et al.~(CCS 2015) have recently demonstrated, these constructions remain vulnerable to several
attacks.
Additionally, all previous ideal OPE schemes (with or without frequency-hiding)
either require a large round complexity of $O(\log n)$ rounds {\em for each
insertion}, or a large persistent client storage of size $O(n)$,
where $n$ is the number of items in the database. It is thus desirable to
achieve a range query scheme {\em addressing both issues gracefully}. 

In this paper, we propose an alternative approach to range queries over
encrypted data that is optimized to support insert-heavy
workloads as are common in ``big data'' applications while still maintaining
search functionality and achieving stronger security.  Specifically, we propose
a new primitive called partial order preserving encoding (POPE) that achieves
ideal OPE security {\em with frequency hiding} and also leaves {\em a sizable
    fraction of the data pairwise incomparable}. 
Using only
$O(1)$ persistent and $O(n^\epsilon)$ non-persistent client storage for
$0 < \epsilon < 1$, our POPE scheme provides extremely fast batch insertion
consisting of a single round, and efficient search with $O(1)$ amortized cost
for up to $O(n^{1-\epsilon})$ search queries. 
This improved security and performance makes our scheme better suited for
today's insert-heavy databases.

\end{abstract}

\section{Introduction}\label{sec:intro}

\paragraph{Range queries over big data.}
A common workflow in ``Big Data'' applications is to collect and store a large
volume of information, then later perform some analysis (i.e., queries) over the
stored data.  In many popular NoSQL key-value stores such as Google BigTable
\cite{CDGHWBCFG06} and its descendants,
e.g.~\cite{DHJKLPSVV07,Accumulo,Cassandra,HBASE}, the most important query
operation is a \emph{range query}, which selects rows in a contiguous block
sorted according to any label such as an index, timestamp, or row id. 

In order to support high availability, low cost, and massive scalability, these
databases are increasingly stored on remote and potentially untrusted servers,
driving the need to secure the stored data. While traditional encryption
protects the confidentiality of stored data, it also destroys ordering
information that is necessary for efficient server-side processing, notably
for range queries. An important and practical goal is therefore to provide data
security for the client while allowing efficient query handling by the database
server.

In many big data scenarios, {\em a moderate number of range queries over a
huge amount of data} are performed.  For example, a typical application might be
the collection of data from low-powered sensor networks as in \cite{WGA06},
where insertions are numerous and happen in real-time, whereas queries are
processed later and on more capable hardware. 
In this work, we target this type of scenario.

\begin{figure*}[t]
\small
\begin{center}
\begin{tabular}{l|c|c|c|c|c|c}
\hline
\multirow{2}{0.5in}{} & \multicolumn{2}{c|}{Comm.\ Rounds} &
Amortized &
\multicolumn{2}{c|}{Client Storage} & Incomparable\\
\cline{2-3}\cline{5-6}
&Insert & Query & Communication & Working set & Persistent & Elements \\
\hline
Here & $1$ & $O(1)$ & $O(1)$ &$O(n^\epsilon)$& $O(1)$&
$\Omega\left(\tfrac {n^{2-\epsilon}}{m} - n\right)  $ \\
Popa et al. \cite{PLZ13} & $O(\log{n})$&$O(\log{n})$&$O(\log{n})$&$O(1)$&$O(1)$&0\\
Kerschbaum \& Schr\"opfer \cite{KS14}& $1$ & $1$ & $O(1)$ & $O(n)$ & $O(n)$&0\\
\hline
\end{tabular}
\end{center}
\caption{
Comparison of OPE-based range search schemes.  $n$ is the total number
of inserts, and $m$ is the total number of search queries. The communication complexity is given in number of encrypted
elements. For our scheme we require at most $O(n^{1-\epsilon})$ total number of
queries. In \cite{PLZ13} and ours, the $O(1)$ persistent storage is used for
storing the encryption key.  The incomparable elements refers to the number of element
pairs (out of $\Theta(n^2)$ total) that cannot be compared even after $m$ queries are performed.  
\label{fig:comp}}
\end{figure*}

\paragraph{Range queries with order-preserving encoding (OPE).}
A simple and efficient solution for performing range queries over encrypted data
was recently proposed by Popa et al.~\cite{PLZ13} who showed how to build an
order-preserving \emph{encoding} (OPE) scheme \footnote{We abuse notation and
    use OPE to refer to both order-preserving encryption and order-preserving
encoding.}, which guarantees that $\enc(x) < \enc(y)$ iff $x < y$, allowing range
queries to be performed directly over encoded values.  Additionally, this scheme
achieves the \emph{ideal} security goal for OPE of IND-OCPA
(indistinguishability under ordered chosen-plaintext attack)~\cite{BCLO09} in
which ciphertexts reveal no additional information beyond the order of the
plaintexts.  This scheme differs from traditional encryption in two ways.
First, the encoding procedure is \emph{interactive} requiring multiple rounds of
communication between the data owner (client) and the database (server).
Second, the ciphertexts produced are \emph{mutable} so previously encoded
ciphertexts may have to be updated when a new value is encoded. This approach requires
$O(\log n)$ rounds of communication and $O(1)$ client storage, where
$n$ is the number of items in the database. 

A different trade-off between client storage and communication is given by
Kerschbaum and Schr{\"o}pfer~\cite{KS14} achieving just $O(1)$ communication
to encode elements (from a uniform random distribution), but requiring $O(n)$
\emph{persistent} client storage to maintain a directory providing the mapping
between each OPE ciphertext to the corresponding plaintext --- proportional to
the storage requirements on the remote database itself.

When used for range searches over encrypted data, these two schemes either require
significant communication, or significant client storage.  Moreover, in the second of these
schemes the directory in the persistent client storage depends on the
full dataset.  Thus
it is not easily amenable to a setting with multiple inserting clients, a common 
deployment scenario in big data applications (e.g., multiple weak sensors
encrypting and inserting data for analysis in the cloud), as the persistent storage has to 
be synchronized across all the clients.

Hence, we ask the following question:

\begin{itemize} 
\item[]
\em 
In the scenario of a large number of insertions and a
moderate number of range queries, can we design a secure
range-query scheme with both \emph{small,
non-persistent client-side
storage} and \emph{much lower communication cost}? 
\end{itemize}

\paragraph{Toward stronger security: frequency-hiding and more.}
As recently pointed out by Naveed et al.~\cite{NKW15}, security provided by OPE
may be insufficient for many applications.  Specifically, they showed attacks
that use frequency analysis and sorting of ciphertexts to decrypt OPE encrypted
values using some auxiliary information.  To counter the first of these attacks,
Kerschbaum~\cite{K15} proposed a stronger notion of security (IND-FAOCPA) that
also hides the \emph{frequency} of OPE-encoded elements (i.e.  hides equality).
However, even this does not address all known attacks on OPE.
Hence, this paper asks the following question:
\begin{itemize} 
\item[]
\em 
Can we design an efficient range query scheme with security \emph{better
than frequency-hiding}?
\end{itemize}

\subsection{Our Work}

\paragraph{Our contribution.}
In this paper we give a positive answer to both of the above questions,
proposing an alternative range query scheme that we call partial order
preserving encoding or POPE.  Specifically, our POPE construction satisfies the
following properties when storing $n$ items using $O(1)$ persistent and
$O(n^\epsilon)$ working storage on the client and performing at most $O(n^{1-\epsilon})$
range queries for any constant $0 < \epsilon < 1$:

\begin{itemize}

\item Trivial insert operations consisting of 1 message from the client to the
    server and no computation for the server. Furthermore, a large number of
    data insertions can be performed only with a single round in a batch. 

\item $O(1)$-round (amortized) communication per range query.

\item No persistent client storage between operations except the encryption key.

\item Greater security than IND-FAOCPA. Our scheme leaks nothing beyond the
    order of (some of the) inserted elements while also hiding equality.
    Moreover, a fraction of plaintext pairs remain incomparable even after the
    queries.

\end{itemize}

See Figure~\ref{fig:comp} for how this compares to existing schemes.

We have implemented our construction and tested it on a variety of workloads,
comparing to other schemes and also measuring its network performance.
We find that our scheme is especially suitable for typical big data applications
where there are many more inserts than queries. As an example data point, with
about one million insertions and one thousand range queries, our POPE scheme is
20X faster than the scheme by Popa et al.  

\begin{figure*}[t]
\includegraphics[width=\textwidth]{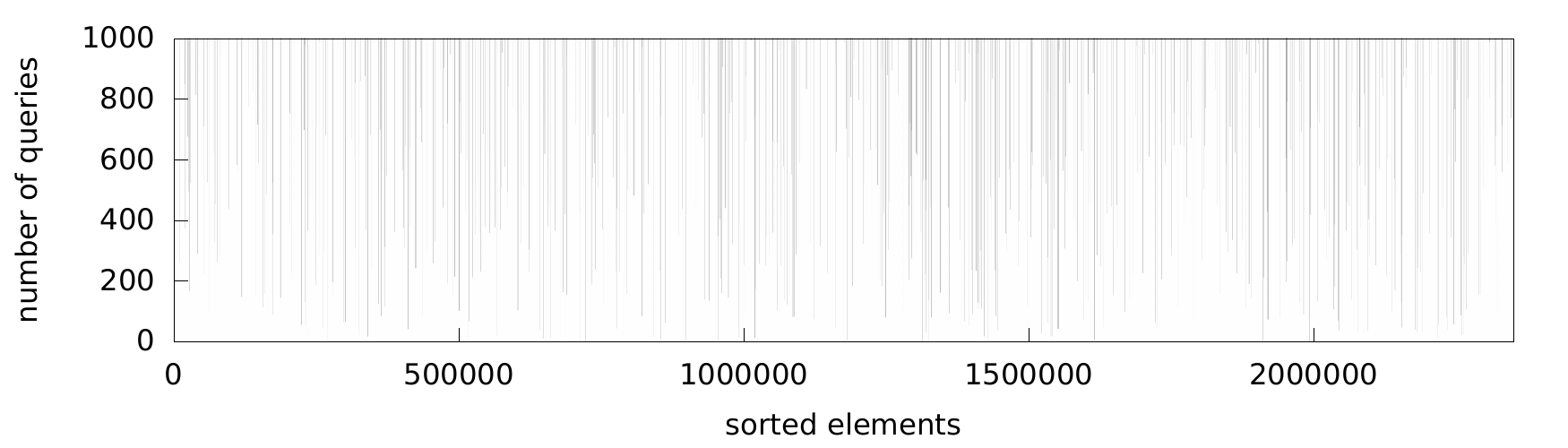}
\caption{Server's view of salary data whose order is incrementally revealed
after inserting more than 2 million salary entries and then performing
1000 random range queries.  The black lines indicate entries whose order
is known by the server, while the white regions indicate entries that remain
pairwise incomparable after some number of queries.\label{fig:incomp}}
\end{figure*}

We also experimentally validate our claim of improved security by observing how
many data items remain unsorted (i.e., the server cannot learn their relative order) 
after some number of queries are performed over real-world data. Specifically, 
we ran an experiment where we inserted over 2 million
public employee salary figures from \cite{caldat14} and then performed 1000
random range queries.  \cref{fig:incomp} shows the server's view of the salary data 
after various numbers of queries.  The black lines indicate elements whose position in the 
global order the server knows (the shading of the lines indicates the fraction of comparable points in each value range with lighter shading indicating a lower fraction), while the contiguous white regions represent data points whose relative
order is unknown.  Note that for a typical OPE scheme, this image would be fully black (all order revealed).

See Section~\ref{sec:evaluation} for more details on our implementation
and further experimental data.

\paragraph{POPE tree: no sorting when inserting data.}
Our main technique to make this possible is {\em lazy indexing}.  Specifically,
unlike OPE, {\em we do not sort the encoded values on insert, instead only
partially sorting values when necessary during query execution}. If we regard the
actual location in the search tree data structure as an implicit encoding of an
encrypted value, our scheme gives a partially ordered encoding, and hence the
name of our construction, POPE (partial order preserving encoding).

In particular, our scheme works by building a {\em novel} tree data structure
(inspired by buffer trees~\cite{Arg03}), which we call a POPE tree, where every
node contains an {\em unsorted buffer} and a sorted list of elements.  The
invariant that we maintain is that the sorted elements of a node impose a
partial order on both sorted and unsorted elements of its child nodes.  That is,
all sorted and unsorted values at child $i$ will lie between values $i-1$ and
$i$ in the parent's sorted list.  We stress that there is no required relation
between unsorted elements of a node and the elements of its child nodes. In
particular, unsorted elements of the root node do not need to satisfy any
condition. That said, one can simply insert a value by putting it in the
unsorted buffer of the root node. 

Having the server incrementally refine this POPE tree on range queries allows
us to achieve both better efficiency and stronger security. In particular, 
\begin{itemize}
\item Insertion is extremely simple by putting the encrypted label in the
    unsorted buffer of the root of the POPE tree. Moreover, a large number of
    items can be inserted in a batch, and the entire task takes only a single
    round. We note that the interactive OPE scheme in~\cite{PLZ13} cannot
    support a batch insertion, since each insertion is involved with traversing
    and changing the encoding tree structure, and it's quite difficult to
    parallelize this procedure maintaining the consistency of the tree
    structure. 

\item The cost of sorting encrypted elements can be amortized over the queries
    performed. In particular, {\em on each query we only need to sort the part
    of the data that is accessed during the search, leaving much of the data
untouched.} This allows us to support range queries with much better efficiency
and simultaneously achieve stronger security by having some fraction of
pairs of elements remain incomparable. 
    
\item Since encodings are sorted during searches, the cost of performing a batch
    of search queries is often much cheaper than performing these queries
    individually, as later queries do not need to sort any elements already
    sorted in earlier queries.  
\end{itemize}

We now describe the key properties of our data structure in more detail. Intuitively, thanks to the
required condition between sorted elements of a node and the elements of its
child nodes, the sorted values at each node can serve as an array of
simultaneous pivot elements for the elements in the child nodes, in the sense of
Quicksort~\cite{quicksort}.  To maintain this property we make use of client
working-set storage to partition a set of unsorted elements according to the values
at the parent.  Specifically, we require the client to be able to read in a
list of $O(n^\epsilon)$ encrypted values and then to partition a
stream of other encrypted values according to
these split points.  Using this amount of client working-set storage we can ensure
that the depth of the buffer tree remains $O(1)$, allowing for low amortized
latency per client query.  Note that any elements stored in the same unsorted
buffer at the end of the procedure remain incomparable.

\section{Preliminaries}

\subsection{Security with No Search Queries} \label{ssec:sec_noqs}
The security definitions of OPE variants consider how much information is
revealed by the ciphertexts that are created when data is inserted. This
measure is important since OPE ciphertexts must inherently reveal ordering
information. The ideal security achievable {\em even without any search
queries} is revealing ordering information of the underlying plaintexts but
nothing more. Our POPE scheme, however, gives much stronger guarantee of
revealing no information about the underlying plaintexts during insertion.
Instead,  ordering information is {\em gradually leaked} as more and more
search queries are performed. In this section, we briefly discuss the security
guarantees that OPE variants and our scheme provide, before any search queries
are performed. 

\paragraph{Security of OPE.}
The security notion for OPE schemes is IND-OCPA (indistinguishability under
ordered chosen-plaintext attack) \cite{BCLO09,PLZ13}: Ciphertexts reveal no
additional information beyond the order of the plaintexts.  However, Naveed
et~al.~\cite{NKW15} demonstrated this level of security is sometimes
insufficient, by showing how the revealed order can be used to statistically
recover a significant amount of plaintext data in an OPE-protected medical
database.

\paragraph{Security of frequency-hiding OPE.}
To address the above issue, Kerschbaum~\cite{K15} proposed a stronger security
notion, called \emph{indistinguishability under frequency-analyzing
ordered chosen plaintext attack} (IND-FAOCPA). Informally, the definition
requires that ciphertexts reveal no additional information beyond {\em a
randomized order} of the plaintexts.  A \emph{randomized order} $Y$ (some
permutation of $[n]$ for $n$-element sequences) of some sequence $X$ of possibly
non-distinct elements is an ordering you can obtain from the sequence by
breaking ties randomly.
For example, the randomized order of $X_1 = (1, 4, 2, 9)$ or $X_2 = (2, 8, 5,
20)$ could only be $Y_1 = (1, 3, 2, 4)$ (meaning ``first in sorted order was
inserted first, third in sorted order was inserted next,'' and so on)
because $X_1, X_2$ began totally ordered.  However, the sequence $X_3 = (1, 2,
2, 3)$ has two possible randomized orders, namely $Y_2 = (1, 2, 3, 4)$ and $Y_3
= (1, 3, 2, 4)$. 

Note that for any randomized order, e.g. $Y_1 = (1, 3, 2, 4)$, there are
many sequences that could map onto it (depending only on the domain of
the sequence and the constraints imposed by known partial order information
on the sequence). 
This property of a randomized order is useful for hiding
\emph{frequency}.  The
motivating example for frequency-hiding security is a database that stores a
large number $n$ of encodings for which the underlying label space $\mathcal{L}
= \{\ell_1, ..., \ell_T\}$ is small, i.e., $T \ll n$. For example, \cite{K15}
considered a setting where each label is either $\ell_1$ = ``female'' $(F)$ or
$\ell_2$ = ``male'' $(M),$ with the sequence $(F, F, M, M)$ ideally encoded as,
say, $(2, 1, 3, 4)$. Examining only $(2, 1, 3, 4)$ does not reveal if the
underlying sequence was originally $(F, F, F, F), (F, F, F, M), 
(F, F,\allowbreak M,  M), (M, F, M, M),$ or $(M, M, M, M)$.

To turn an OPE scheme into a frequency-hiding OPE scheme, consider adding a small, random
fractional component to the OPE-ordered field during encoding, e.g. $X_1 = (1,
1, 2, 2)$ becomes e.g. $X'_1 = (1.12, 1.36, 2.41, \allowbreak 2.30)$,  which
\emph{randomly} maps $X_1$ to the ordering $Y = (1, 2, 4, 3)$, and then $X'_1$
is encoded under the OPE scheme. In \cite{K15}, this type of scheme is shown
IND-FAOCPA secure in the \emph{programmable random oracle model}.  

We use this approach to add frequency hiding in POPE.  
However, even this stronger definition fails to protect against all known attacks as it still 
reveals the order between all distinct plaintexts in the database, allowing 
for sorting-based attacks. 

\paragraph{Security of POPE.}
POPE, on the other hand, fully hides all inserted plaintexts until search
queries are performed.  Looking ahead, our POPE scheme encrypts an item using
semantically secure encryption. This implies in the POPE construction, {\em
ciphertexts reveal no information} about the underlying plaintexts. Of course,
it is not sufficient to just discuss security on insert without considering what
happens on queries.  Thus, we give a security definition below capturing what
happens both on insertion and during search queries. 

\subsection{Security with Search Queries}
We propose a simulation-based definition that captures both ideal OPE security and frequency-hiding
even when considering what happens during the search procedure.  Specifically,
we require the existence of a simulator simulating the view of the protocol execution 
given only a randomized order of (some of) the plaintexts.  We model this by a
\emph{random order} oracle $rord$ which just takes the indices of two data items and 
returns which item is larger according to some fixed randomized order.  Since the simulated 
view is constructed using only this oracle,
the only information leaked in the real protocol corresponds to the oracle queries made to the $rord$
oracle, i.e., the randomized order on the queried plaintexts.

To formalize the simulation tasks, for a sequence $\seq$ of insertion and search
operations, we define the profile $\profile(\seq)$ of sequence $\seq$ to be a
sequence where each value in $\seq$ is replaced with a unique index (simply
incrementing starting from 1) to identify the operation. An example sequence and
its profile can be:
\hspace{-1em}\begin{eqnarray*}
&& \seq: (\ins~ 10, \ins~ 100, \range~ [8, 20], \ins~ 41). \\
&& \profile(\seq): (\ins~ 1, \ins~ 2, \range ~ [3,4], \ins~ 5). \\
\end{eqnarray*}

\vspace{-2em}
\BD \label{def:sec-pope}
A range query protocol $\Pi$ is called frequency-hiding order-preserving, for
any honest-but-curious server $S$, if there
is a simulator $\Sim$ such that for any sequence $\seq$ of insertions and searches,
the following two distributions are computationally indistinguishable:
$$\view_{\Pi,S}(\seq) ~\approx_c~ \Sim_\Pi^{rord_\seq(\cdot,
\cdot)}(\profile(\seq)), $$ where the left-hand side denotes the real view of
$S$ when executing the protocol $\Pi$ with $\seq$ as the
client's input, and the right-hand side is the output of the simulator $\Sim$
taking as input $\profile(\seq)$ and referring to oracle $rord$.  The oracle
$rord_{\seq}(\cdot, \cdot)$ works as follows:
\BI
\item[] \hspace{-1em} $rord_\seq(i, j)$: It is initialized with a randomized order
$\pi$ of the labels in $\seq$ by breaking ties randomly. Then, for each query $(i,
j)$, return whether the $i$th label has a higher ranking than the $j$th,
according to $\pi$. 
\EI
\ED

Since the simulator refers to only the profile and the oracle, we can say that
for any protocol satisfying the above definition, the protocol transcript leaks
to the server only the profile and the randomized order of the queried
plaintexts. 
One benefit of this definition is it covers both non-interactive FH-OPE schemes
and our interactive POPE scheme. 

\paragraph{Leaking only a partial order.} Recall that our POPE scheme gradually
leaks the ordering information as more comparisons are made in order to execute
the queries.  To formally treat the amount of information that remains hidden
after some number of range queries, we introduce a definition that captures the
number of points that remain incomparable even after some queries are performed.  

First, we explain what we mean by the number of incomparable element pairs {\em
with transitivity}.  For example, consider a sequence of four labels $\labels$ =
($\ell_1$, $\ell_2$, $\ell_3$, $\ell_4$). There are ${4 \choose 2}  = 6$
initially unordered pairs: $\{\ell_1, \ell_2\}$, $\{\ell_1, \ell_3\}$,
$\{\ell_1, \ell_4\}$, $\{\ell_2, \ell_3\}$, $\{\ell_2, \ell_4\}$, $\{\ell_3,
\ell_4\}$.  During query execution the order of some of these pairs
may become known to the server, i.e., if it queries the $rord$ oracle on the indices
of some such pair or if the order can be inferred from its previous queries.
For example, given $\info$ =
($\ell_1 > \ell_2$, $\ell_2 > \ell_4$), then due to transitivity, the server can infer
$\ell_1 > \ell_4$.  However, the following pairs still remain {\em incomparable}: 
$$\{\ell_1, \ell_3\}, \{\ell_2, \ell_3\}, \{\ell_3, \ell_4\}$$ 
Armed with this notion of incomparable pairs with transitivity, we give the
following definition:

\BD
Let $n, m$ denote the number of insertions and range
searches respectively. A range query protocol $\Pi$ is frequency-hiding partial order
preserving with \emph{$u$ incomparable element pairs with transitivity},
if for any operation sequence $\seq$ with $n$ inserts and $m$ range
queries, the simulator successfully creates a simulated view required by
Definition~\ref{def:sec-pope} while leaving at least $u$ pairs of elements that
are incomparable with transitivity based on the queries made by the simulator to $rord$.
\ED

In this paper, whenever we consider incomparable pairs, we consider it with
transitivity, and from now on, we will omit the phrase ``with
transitivity''. 
Note that both the OPE scheme by Popa et al.~\cite{PLZ13} and the FH-OPE scheme by Kerschbaum~\cite{K15} have $0$ incomparable element pairs for any $n$
inserts, even with $0$ searches. However, our POPE scheme shows a more gradual
information leakage. We discuss this in more detail in
Section~\ref{sec:analysis}.

\section{Main Construction}\label{sec:construction}

\subsection{Overview}
Our scheme consists of a client and a server, which we denote by \cl and
\ser respectively. \cl holds an encryption key and performs
insertions and range query operations through interactive protocols with
\ser. (In fact, only the range query operation is interactive, which
is a key benefit of our construction!)

As \cl is \emph{stateless} and needs to remember nothing (other
than the secret key), all data is stored encrypted by \ser. To organize
this data and facilitate fast lookups, \ser maintains a POPE tree to
hold the ciphertexts. The high-level structure of this tree is similar
to a B-tree, where each node has a bounded number of children and all
leaf nodes are at the same depth. In fact, the number of children of any
POPE tree internal node is between $L/2+1$ and $L+1$, where $L$ is the
local temporary storage capacity of \cl.

Where the POPE tree differs from a standard B-tree is that every node
contains an unsorted \emph{buffer} of ciphertexts \emph{with unbounded
size}. The benefits of our construction, both in terms of efficiency and
security, stem from the use of these unsorted buffers. For efficiency,
they allow to delay expensive sorting and data movement operations until
necessary to execute a range query. Security benefits stem from
the fact that the relative order of elements in the same unsorted buffer
is not revealed to an attacker.

The \textbf{insertion} protocol is trivial: \cl encrypts the plaintext
value to be inserted and sends it to \ser, who simply appends the new
ciphertext to the root node's unsorted buffer. 
Because semantically secure encryption is used, the ciphertexts do not reveal
anything about their true values or order, not even whether two inserted
values are the same or different.
All of the actual sorting
and ordering is delayed until queries are performed. 

Before completing a range query, \ser interacts with \cl to
\textbf{split} the tree according to each of the two query endpoints.
This subroutine --- the most sophisticated in our entire construction
--- has three stages. First, for all the internal POPE tree nodes along
the search path for the query endpoint, the unsorted buffers are cleared
out. This clearing of the buffers proceeds from root to leaf, and
involves streaming all buffer ciphertexts back from \ser to \cl, who
responds for each one with the index of which child node that ciphertext
should flow down to. Recall that we maintain each internal node having
at most $L+1$ children; this allows the operation to be performed
efficiently by \cl without overflowing the client's size-$L$ local
storage.

This initial stage of the \textbf{split} ends at a leaf node. The second
stage involves reducing the size of that leaf node's buffer to at
most $L$, the size of \cl's local storage. This leaf node buffer
reduction proceeds by selecting $L$ random ciphertexts from
the leaf node's buffer, and using \cl to split the single leaf into $L+1$ new
sibling leaf nodes, according to these randomly-selected elements. These
$L$ randomly sampled ciphertexts are inserted into the parent node as
partition elements between the new leaf nodes.  This leaf node splitting procedure
is repeated until the resulting leaf node has a buffer of size at most $L$.

However, we may have inserted too many new children into the parent
node, causing it to have more than the limit of $L+1$ children. So a
\textbf{rebalance} operation must finally be completed, from the leaf back
up to the root node, creating new internal nodes as necessary until they
all have at most $L+1$ children as required. Note that this stage does
not require any further ordering or consultation with \cl.

After performing this split operation for both endpoints, the actual
\textbf{range query} can now be completed by
\ser returning to \cl all the ciphertexts in all buffers of nodes between the
two query endpoints. Again, this does not require any further ordering
information from \cl. 
Of particular importance for security is that there may be large
unsorted buffers even after the range query completes, because all
contents of those buffers lie entirely within or outside of the desired
range. The server either returns all of none of the ciphertexts in these
buffers, but still does not (and does not need to) learn their order.

\paragraph{Parameters.}
Recall that the parameter $n$ represents the total number of items inserted into the
database, and the parameter $m$ represents the total number of range query operations
performed. The client can temporarily store $L +
O(1)$ labels in its local memory for the duration of a given query.  Let $L = n^\epsilon$ for constant $0<\epsilon <1$.

\paragraph{Notation.} 
To support realistic application scenarios, we distinguish between two types
of data that \ser holds: {\bf (i)} \emph{labels} $\ell$ and {\bf (ii)} \emph{blocks}
that are composed of a POPE-encoded label $\ell$ and an arbitrary, encrypted payload $v_\ell$.
This models the case when range searches over POPE-encoded labels are used to retrieve
the payloads.  No searching directly over the payloads is supported.

We remark that, in principle, for every distinct label $\ell$, there could be many
distinct blocks $(\ell, v_{\ell_1}), (\ell, v_{\ell_2}), ...$ stored by \ser.
However, we will
restrict to the special case when for each label $\ell$ there is at most one
block $(\ell, v_\ell)$ in order to convey the main ideas more
clearly. (Note this distinctness property holds w.h.p.\ if we use the tie-breaking 
procedure described in Section~\ref{ssec:sec_noqs}.)

\subsection{Encryption of Labels}
In our system, whenever \cl\ communicates a label $\ell$ to  \ser\ we have \cl 
always send a ciphertext $\bar \ell$ to \ser, where $\bar \ell \from \Enc_k(\ell)$. 
Besides an encryption of the label itself, this ciphertext must also
encrypt 
(a) the tie-breaking random value necessary for
frequency-hiding POPE and (b) an indication of the label's \emph{origin} (left or right
query endpoint, or insertion).

\paragraph{Tie-breaking randomness.}
Consider for example
that the labels $(1,2,2,3)$ have been inserted, followed by a range
query for all values between $2$ and $3$, requiring a total of six
encryptions.
From \cref{ssec:sec_noqs},
tie-breaking randomness can be thought of as adding a random
fractional part to each plaintext before encrypting, so for example we
encrypt the labels $(1.89, 2.15, 2.35, 3.93)$ and the range query
endpoints $2.23$ and $3.38$.

\paragraph{Origin bits.}
This hides the repeated label $2$, but creates a new problem:
the labels $2.15$ and $3.93$ which should be
included in a range search between 2 and 3, would be excluded because of
the tie-breaking. So we also include two bits $\pi$ for the \emph{origin} of
the plaintext: $\pi_l=00$  and $\pi_r=11$ for left and right  query
endpoints respectively, and $\pi_m=01$ for
an insertion. These bits are
inserted between the actual label and the tie-breaking values, so
(continuing the previous example), we would insert the encryptions of
$(1.01.89, 2.01.15, 2.01.35, 3.01.93)$ and query endpoints
$2.00.23$ and $3.11.38$. This forces the range search to return the
three correct values.

\paragraph{Two-block ciphertexts.}
Even treating the two origin bits as part of the label, each
plaintext becomes two blocks long, so that a straightforward application
of CTR or CBC mode encryption results in ciphertexts of three blocks.
One can achieve
better efficiency by not including the tie-breaking randomness but still
enabling the receiver to compute it.  
In particular, let $f$ be a PRP, and let:
\BI
\item $\enc_k(m\|\pi)$: Choose a random string $r$. Return the pair
  $(r,\  f_k(r+1) \xor (m\|\pi))$.
\item $\dec_k(c_1, c_2)$: Compute $m\|\pi \from f_k(c_1+1) \xor c_2$ and 
   the tie breaking randomness $u \from f_k(c_1+2)$. Return $(m, \pi, u)$, 
\EI

Note it's just the CTR mode of encryption.
Even though the ciphertext doesn't explicitly contain the tie-breaking
randomness, the reconstructed $u$ serves for this purpose. 

\subsection{Server Memory Layout} 

\ser statefully maintains the {\sf POPE~tree~$\mathcal{T}$}, which
is a balanced $L$-ary tree with root $r.$
\BI
\item Each \emph{non-leaf} node $u\in\mathcal{T}$ stores a ${\sf
buffer}$ and a ${\sf list}$. 
\item Each \emph{leaf} node $u\in\mathcal{T}$ stores a ${\sf buffer}$ only.
\EI
A ${\sf buffer}$ stores an unbounded, \emph{unsorted} set of (encryptions of) blocks 
$\{(\ell_1,v_{\ell_1}), (\ell_2,v_{\ell_2}), ..\}$, and
a ${\sf list}$ stores at most $L$ \emph{sorted} (encryptions of) labels $(\ell_1, ..., \ell_L)$.

\paragraph{Main invariant of the POPE tree $\mathcal{T}.$} 
We will enforce the following, main order-invariant on \ser's tree $\mathcal{T}$:
\begin{center}
\noindent
Let $\ell_{j-1}$ and $\ell_j$ be the $(j-1)$th and $j$th sorted labels at some
(non-leaf) node $u$ in $\mathcal{T}$. Then, for all labels $\ell$ in the sub-tree
$\mathcal{T}_{u_j}$ rooted at the $j$th child $u_j$ of $u$, we have $\ell_{j-1}
< \ell \le \ell_j$.
\end{center}
Intuitively, this guarantee of global partial ordering enables the $L$ sorted
labels $\ell_1, ..., \ell_L$ at each node $u$ to serve as an array of
simultaneous \emph{pivot elements}, in the sense of Quicksort~\cite{quicksort},
for the $L+1$ sub-trees rooted at $u$'s (at most) $L+1$ children $u_1, ...,
u_{L+1}$. Looking ahead, we use this simple, parallel pivot idea in conjunction with
the parameter setting $L=n^\epsilon$,
implying $\mathcal{T}$ has depth $\lceil{1/\epsilon}\rceil = O(1)$, to enable
\ser to traverse and maintain the tree $\mathcal{T}$ with low amortized
latency over repeated batches of \cl queries.

\subsection{The POPE Protocol}
\label{sec:protocol}

We now present more formally our protocol POPE consisting of three
operations: \Psetup, \Pinsert, and \Psearch. The
\Psearch protocol results in additional calls to helper protocols
\Psplit and $\mathsf{Rebalance}$, described afterward.

\filbreak

\paragraph{Implementing Setup.} At \Psetup, \cl and \ser do:
\begin{framed}
\small
\noindent\underline{\Psetup}:\medskip

\noindent{--} \cl generates private keys for label/block encryption.

\noindent{--} \ser initializes $\mathcal{T}$ as a root $r$ with empty buffer and list.
\end{framed}

\paragraph{Implementing Insert.} To \Pinsert a \emph{block}
$(\ell,v)$, \cl and \ser do:
\begin{framed}
\small
\noindent\underline{\Pinsert$(\ell,v)$}:\medskip

\noindent{--} \cl sends (encrypted) block $(\ell,v)$ to \ser.

\noindent{--} \ser appends block $(\ell,v)$ to the end of the current root
node's  buffer.
\end{framed}

\begin{figure}[!ht]
\begin{center}
\includegraphics[width=.5\linewidth]{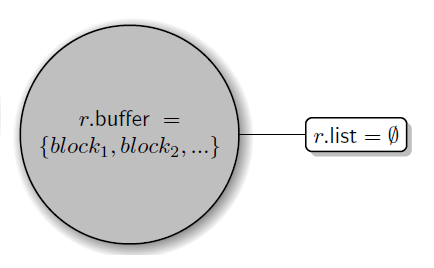} %
\caption{The state of \ser's tree $\mathcal{T}$ prior to any \Psearch queries.}\label{fig:split}
\end{center}
\end{figure}

After \Psetup and possibly many \Pinsert operations (but no \Psearch operations), the POPE tree $\mathcal{T}$ 
held by \ser appears as in Figure~\ref{fig:split}.

\paragraph{Implementing Search.} For \cl to \Psearch for the range of blocks held by \ser in $\mathcal{T}$ between two \emph{labels} $\ellL$ and $\ellR$, \cl and \ser do:

\begin{framed}
\small
\noindent\underline{\Psearch$(\ellL, \ellR)$}:\medskip

\noindent{--} \cl and \ser engage in an interactive protocol \Psplit \emph{twice},\\
\indent once for $\ellL$ and once for $\ellR$.

\noindent{--} After each \Psplit, \cl identifies for \ser the leaf node\\
\indent $u_\mathsf{left}$ (or $u_\mathsf{right}$) in $\mathcal{T}$ that matches the label $\ellL$ (or $\ellR$).

\noindent{--} \ser sends the blocks in $[u_\mathsf{left}, u_\mathsf{right}]$ to \cl.
\end{framed}

\paragraph{How to Split the POPE Tree.} For \cl to \Psplit\ \ser's tree $\mathcal{T}$ at label $\ell\in\{\ellL, \ellR\}$,
\cl and \ser engage in an interactive protocol. This operation will
return the leaf node whose buffer contains the given label with the
guarantee that all nodes along the path from the root to that leaf
have empty buffers.

Individual \Psplit calls always begin at the current root $r\in\mathcal{T}$.
After any (non-leaf) node $u\in\mathcal{T}$ is split, \ser learns (from \cl) the index $i\in [L+1]$ of the next child $u_i$ of $u$ to be \Psplit.
The \Psplit protocol proceeds \emph{recursively} down some path of $\mathcal{T},$ splitting subsequent children $u_i, u_{i, j}, ...$ until terminating at a leaf node $u$.
(For readability in what follows, we assume that \ser always returns
whole nodes to \cl for each \Psearch response.)

We break our description of \Psplit into two broad cases: {\bf (i)} the
{\Psplit}s of \emph{internal, i.e., non-leaf nodes}, and
{\bf (ii)} the {\Psplit}s of \emph{leaf nodes}.

\paragraph{Case (i) --- Splits at internal nodes:} 
For splits at internal nodes $u$ with children denoted $u_i$, \cl and \ser do:

\begin{framed}
\small
\noindent\underline{\Psplit$(\ell)$}\emph{ --- for internal nodes $u$}:\medskip

\noindent{--} \ser sends $\mathcal{L} = u.\mathsf{list}$ to \cl.

\noindent{--} \ser\ {\bf streams} $(\ell',v')\in u.\mathsf{buffer}$ to \cl.

\indent{--} \cl sends the \emph{sorted} index $i\in [L+1]$\\
\indent\;\; of each $(\ell',v')$ in $\mathcal{L}$ to \ser.

\indent{--} \ser appends \emph{block} $(\ell',v')$ to $u_i.\mathsf{buffer}$
\end{framed}

During this operation, 
\cl either {\bf (a)} sees the searched-for label $\ell\in\{\ellL,
\ellR\}$ (and discovers node $u_i$ to proceed to), or {\bf (b)}
discovers the node $u_i$ that may contain label $\ell$ based on its boundary values.

\begin{figure}[!ht]
\begin{center}
\includegraphics[width=\linewidth]{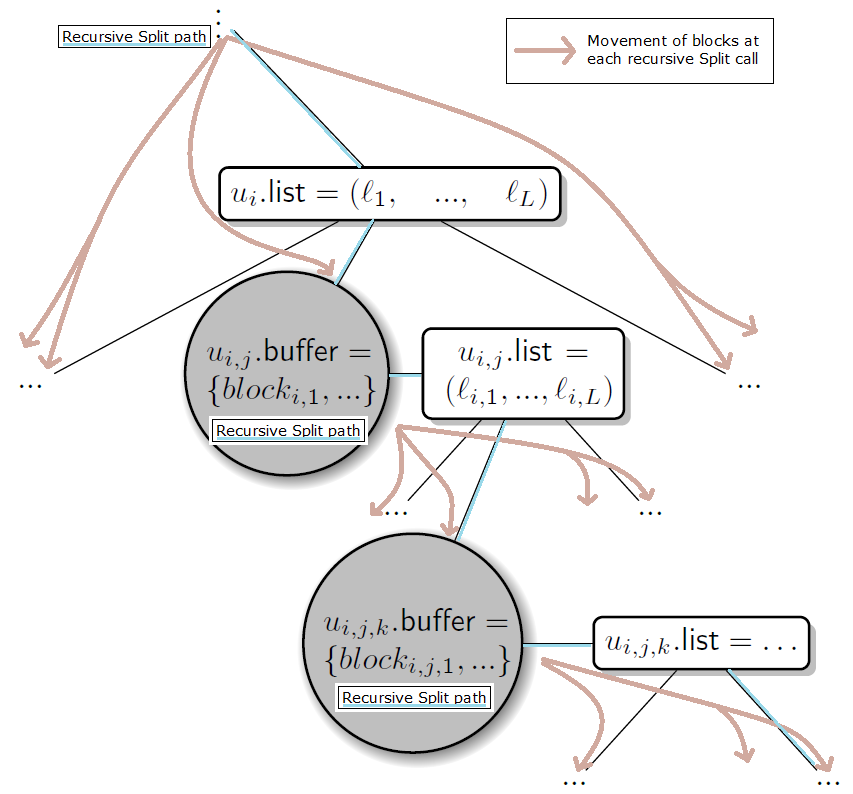} %
\caption{The flow of blocks in recursive {\Psplit}'s of \ser's tree $\mathcal{T}$.}\label{fig:split3}
\end{center}
\end{figure}

The block movement in splits at internal nodes is illustrated in Figure~\ref{fig:split3}. (The outcomes of three ``splits'' are shown.)

\paragraph{Case (ii) --- Splits at the leaves:}
For splits at leaf node $u$ with parent node $u^*$, \cl
and \ser do:

\begin{framed}
\small
\noindent\underline{\Psplit$(\ell)$}\emph{ --- for leaf nodes $u$}:\medskip

\noindent{--} If $|u.\mathsf{buffer}| \le L$, {\bf return}.

\noindent{--} \ser samples $L$ labels $\mathcal{L} = \{\ell_1, ..., \ell_L\}$ from $u.\mathsf{buffer}$.

\noindent{--} \ser creates new root $u^*$ if $u$ is the root node, or
sets $u^*$ to\\
\indent $u$'s parent otherwise.

\noindent{--} \ser sends $\mathcal{L}$ to \cl.

\noindent{--} \cl sorts $\mathcal{L}$ and returns it
to \ser.

\noindent{--} \ser inserts $L$ new sibling leaf nodes $u_i$ into parent $u^*$\\
\indent as well as new labels $\mathcal{L}$ into $u^*.\mathsf{list}$ at the position previously\\ 
\indent occupied by $u$ (node $u$ is deleted after it is split).

\noindent{--} \ser\ {\bf streams} $(\ell',v') \in u.\mathsf{buffer}$ to \cl

\indent{--} \cl sends the \emph{sorted} index $i\in [L+1]$\\
\indent\;\; of each $(\ell',v')$ in $u.\mathsf{buffer}$ to \ser.

\indent{--} \ser inserts \emph{block} $(\ell',v')$ into sibling node $u_i$

\noindent{--} \cl indicates to \ser the index $i$ of new leaf node matching $l$
\end{framed}

Note that if the size of the buffer is smaller than the local storage
capacity $L$ of \cl, then this operation does nothing, and the split is
complete. Otherwise, as in Case (i) of \Psplit, \cl will learn which of the sibling
leaf nodes $u_i$ to recursively \Psplit in order to find label $\ell$.
In this way, a single \Psplit operation may recursively result in
multiple leaf node \Psplit's, with smaller and smaller buffers.

\begin{figure}[!ht]
\begin{center}
\includegraphics[width=\linewidth]{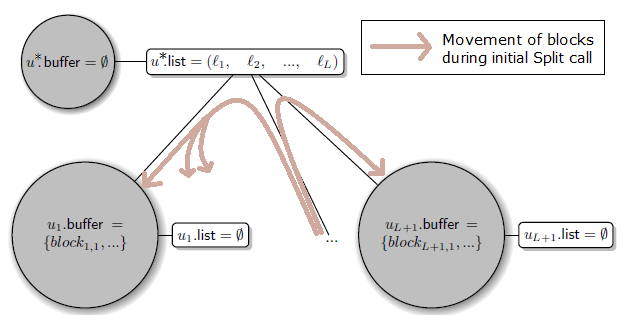}
\caption{The state of \ser's tree $\mathcal{T}$ after the \emph{very first} \Psplit ends: the new root $r := u^*$ (empty buffer, full list), plus $L+1$ leaves.}\label{fig:split2}
\end{center}
\end{figure}

As an example, the new state of \ser's tree $\mathcal{T}$ immediately after \cl's \emph{first} $\mathsf{Split}$ call
(which splits the starting leaf node of $\mathcal{T}$, i.e. the root)
is as depicted in Figure~\ref{fig:split2}.

\paragraph{Clean-up Step: Rebalancing a Split POPE Tree.} 
After completing the \Psplit protocol above, the resulting leaf node at which $\Psplit$ terminates will have
size at most $L$, but some internal node's sorted list may be larger than $L$
because of the insertions from their children --- see case (ii) of \Psplit. This
would be problematic in future \Psplit operations on those internal nodes, as
they would send $u.\mathsf{list}$ to \cl, who only has room for $L$ items.

To fix this, after completing the \Psplit protocol, \ser calls the following
operation on the parent of the resulting leaf node in order to rebalance the
labels in the lists of the internal nodes. We emphasize that
$\mathsf{Rebalance}$ is purely a local data structure manipulation, and does not
require interaction from \cl, since the unsorted buffer of the rebalanced nodes
is empty due to prior \Psplit, having only sorted labels in the list. 

More concretely, to $\mathsf{Rebalance}$ at node $u$ (initially the
parent of the leaf where \Psplit ended),
\ser does:
\begin{framed}
\small
\noindent\underline{$\mathsf{Rebalance}(u)$}:\medskip

\noindent{--} If $|u.\mathsf{list}| \le L,$ {\bf return}.

\noindent{--} If $u$ has no parent $u^*$, create a fresh root node $r$ for $\mathcal{T}$\\ and set $u^* := r$.

\noindent{--} Partition $u.\mathsf{list}$ into sorted sublists of size
at most $L$ each by selecting $\mathcal{L} = [$every $(L+1)$'th element
in $u.\mathsf{list}]$.

\noindent{--} Create $|\mathcal{L}|$ new sibling nodes and insert them as
well as the new labels $\mathcal{L}$ into parent node $u^*$.

\noindent{--} Call $\mathsf{Rebalance}(u^*)$.
\end{framed}

This completes the description of our main POPE protocol.

\section{Analysis}\label{sec:analysis}

\subsection{Cost Analysis} 
We analyze amortized costs on the round complexity and bandwidth per
operation. 

\begin{theorem} \label{thm:cost}
After $n$ insertions and $m$ query operations with local storage of size $L$, our scheme
has the following costs:
\begin{enumerate}
\item $\mathsf{Insert}$ always requires a single round, and $\mathsf{Search}$
requires $O(\log_L n)$ rounds in expectation.  

\item The total expected bandwidth over all $(n+m)$ operations  (excluding the bandwidth
    necessary for sending the search results) is 
    $$O\big(mL\log_L n + n\log_L m  + n\log_L(\lg n) \big).$$
\end{enumerate}
\end{theorem}
The proof is found in Appendix~\ref{sec:app-analysis}.

\paragraph{Remark.}
With $L=n^\epsilon, 0 < \epsilon < 1$, Theorem~\ref{thm:cost} implies that
$\mathsf{Search}$ takes {\em $O(1)$ rounds in expectation}. Moreover, when $L =
n^\epsilon$ and $m = O(n^{1-\epsilon})$ as well, {\em the amortized bandwidth per
operation becomes $O(1)$}.  This is exactly our target scenario of many
insertions and relatively few searches.

\subsection{Security Analysis}

\BT
The POPE protocol is a frequency-hiding order-preserving range query protocol. 
\ET

\begin{proof}
We show that our POPE scheme satisfies Definition~\ref{def:sec-pope} by showing a
simulator. The simulator is very simple. For each insert, the simulator sends
$\enc_k(0)$; due to semantic security of the underlying encryption, the
simulation is indistinguishable. To simulate search queries, the simulator runs
the adversarial server's algorithm, and during the simulation, when the server
needs to compare two encrypted labels, the simulator simply queries the $rord$
oracle to get the answer. It's obvious that the simulated view is
indistinguishable to the real view of the server.  
\end{proof}

\paragraph{Security with queries.}
Range query schemes leak some information of underlying plaintexts from adaptive
search queries. In this case, one important security measure can be the
\emph{number of pairs of incomparable elements}. In any range query scheme,
search queries reveal some partial order on the underlying plaintexts.  Recall
that a \emph{partial order} $\prec$ on a set of elements $S$ is isomorphic to a
directed acyclic graph, closed under transitive closure, whose nodes are
elements of $S$ and whose edges encode the binary relation. A total order on $n$
items always has $\binom{n}{2}$ edges. In any partial order, two elements $x,y
\in S$ are said to be \emph{incomparable} iff neither $x \prec y$ nor $y \prec
x$. In a total order (such as the \emph{randomized order} of \cite{K15}), no
pair of elements is incomparable.
In our POPE scheme, each search query {\em gradually} leaks the ordering
information of the underlying plaintexts. In particular, with a small number of
search queries, there will be many pairs of incomparable elements.

\BT \label{thm:po}
After $n$ insertions and $m$ query operations with local storage of size $L$, where $mL
\in o(n)$, our POPE scheme is frequency-hiding partial-order-preserving with
$\Omega \Big ( \frac {n^2} {mL\log_L n} - n \Big )$ incomparable pairs of
elements. 
\ET

\begin{proof}
Note the simulator in the above proof uses oracle $rord$ whenever the server
algorithm needs to compare the elements. So, we can prove the theorem by using
a counting argument on the number of labels that the server compares.  We
model the server's view of the ciphertext ordering as some $k$ ciphertexts whose
order is completely known, and where the remaining $n-k$ ciphertexts are
partitioned into one of $k+1$ buckets according to the $k$ ordered ciphertexts.
Essentially, this is a worst-case scenario where all internal node buffers in
the POPE tree are empty, the total size of all internal node sorted lists is
$k$, and the remaining $n-k$ ciphertexts reside in leaf node buffers.

We focus on the round complexity for range queries (insertion gives no change in
the number of comparable elements). From \cref{thm:cost}, the total rounds of
communication for range queries, after $n$ insertions and $m$ range
queries, is $O(m\log_L n)$.  From the construction, {\em each round of
communication can add at most $L$ new ciphertexts to those whose sorted order is
completely known}.

Therefore, in the worst case, the server has $k = O(mL\log_L n)$ ciphertexts in
its sorted order, creating $k+1$ buckets in which the other values are placed.  
Thus, the worst-case split that minimizes the total number of 
incomparable elements is for the remaining values to be partitioned equally among
these buckets.  Thus, we have 
$b = \lfloor (n-k)/(k+1) \rfloor$ ciphertexts in each
unsorted bucket. Each bucket contains $b \choose 2$ incomparable items, for
a total of $(k+1) \cdot {b \choose 2} = \Omega(\frac{n^2}{k} - n )$ incomparable pairs.
\end{proof}

\paragraph{Privacy against a malicious server.}
Note the above theorem considers the worst case. This implies we can easily
achieve privacy against a malicious server with tiny additional costs, that is,
by making sure that (1) all the ciphertexts that the server asks the client to
compare are legitimate, that is, created by the client (to ensure this, the
labels should now be encrypted with IND-CCA2 encryption), and (2) the number of
the server's comparison requests should be within the bounds of
\cref{thm:cost}.

Unfortunately, this augmented system doesn't achieve full malicious security;
in particular, a malicious server may omit some values from the query answers,
although it cannot inject a fake result due to IND-CCA2 security of the
underlying encryption. Efficiently achieving full malicious security is left
as an interesting open problem.

\section{Evaluation}\label{sec:evaluation}

\subsection{Experimental setup}
We have made a proof-of-concept implementation of our POPE scheme in order to
test the practical utility of our new approach. The code is written in Python3
and our tests were performed using a single core on a machine with an Intel Xeon
E5-2440 2.4 GHz CPU and 72GB available RAM.  Our implementation follows the
details presented in Section~\ref{sec:construction}.  The symmetric cipher used
is 128-bit AES, as provided by the PyCrypto library. The full source code of our
implementation is available at \url{https://github.com/dsroche/pope}.

\paragraph{Database size.}
While we performed experiments on a wide range of database sizes and
number of range queries, our ``typical'' starting point is one million
insertions and one thousand range queries. This is the same
scale as recent work in the databases community for supporting range
queries on outsourced data \cite{LLWB14}, and would therefore seem to be
a good comparison point for practical purposes.

\paragraph{Parameters.}
In our experiments, we varied the total database size between one thousand and
100 million entries, each time performing roughly $m=n^{1/2}$ range queries and
with $L=n^{1/4}$ local client storage. That is,
$\epsilon=0.25$ in these experiments.  The size of each range being queried was
randomly selected from a geometric distribution with mean 100; that is, each
range query returned on average 100 results.

\paragraph{Network.}
Our main experiments were performed in a local setup, but with careful
measurement of communication and under the assumption that network bandwidth
(i.e., amortized communication size) and latency (i.e., round complexity) would
be the bottlenecks of a remote cloud implementation. 
In particular, in our network experiments, we used the \texttt{tc}
``traffic control'' utility to add specific latency durations as well as
bandwidth limitations. This allowed us to test the behavior under controlled but
realistic network settings, when we throttled the network slower than 5ms of
latency and 20Mbps bandwidth.

\paragraph{Comparison with Popa et~al.}
We compared our construction experimentally to that of Popa et~al.~\cite{PLZ13},
who had a setting most similar to ours.  Further comparison benchmarks, such as
to \cite{KS14} or even to ORAMs, might provide further insight, and we leave
this as future work.

For a fair comparison to prior work, we also implemented the mOPE scheme of
\cite{PLZ13} in Python3 along with our implementation of POPE.  We followed the
description in their work, using a B-tree with at most 4 items per node to store
the encryptions. To get a fair comparison, we used the same framework as our
POPE experiments, with the client that receives sorting and partitioning
requests from the server. In the case of mOPE, each round of communication
consisted of sending a single B-tree node's worth of ciphertexts, along with one
additional ciphertext to be encoded, and receiving the index of the result
within that sorted order. We acknowledge that our implementation is likely less
tuned for efficiency than that of the original authors, but it gives a fair
comparison to our own implementation of POPE.  It is also
important to note that {\em our communication cost measurements depend only on the
algorithm and not on the efficiency of the implementation}.

\paragraph{Measuring communication and running time.}
When our tests measured communication (in terms of rounds and total ciphertexts
transferred) and running time, we did not include the cost
of the server sending the
search results;  this is inherent in the operation being performed and would be
the same for any alternative implementation.

\subsection{Experimental workloads}

\paragraph{Local setting: huge data, various search patterns.}
In our main experiments, where we wanted to scale the number of database entries
from one thousand up to 100 million entries, we used synthetic data
consisting of random pairs of English words for both label and payload values.
For these experiments we also did not actually transfer the data over a network,
but merely measured the theoretical communication cost. This allowed us to test
a much wider range of experimental sizes, as we found a roughly 10x slowdown in
performance when running over a network, even with no throttling applied.

The actual size of each range being searched was, on average, 100 database
entries. While the distribution of searches does not affect the running time of
mOPE, for POPE we varied among three distributions of the random range queries:
(i) uniformly distributed queries, (ii) search queries all ``bunched'' at the
end after all insertions, (iii) a single, repeated query, performed at random
intervals among the insertions. 
According to our theoretical analysis, the ``bunched'' distribution should be
the worst-case scenario and the repeated query should be the best-case.
In practice we did not see much difference in
performance between bunched or random queries, though as expected, we observed
improved performance for the repeated query case.

\paragraph{Networked setting: real salary data.}
To test performance over a realistic network with latency and bandwidth restrictions, we used the California
public employee payroll data from 2014, available from \cite{caldat14}, as a
real dataset on which to perform additional experiments. This dataset lists payroll
information for roughly 2.3 million public employees. We used the total salary
field as our ``label'' value (on which range queries are performed), and the
names as the payload values.

We were not able to complete any test runs of the mOPE using actual network
communication over the salary dataset; based on partial progress in our
experiment we estimate it would take several days to complete just one test run
of this experiment using mOPE and actual network communication with our Python
implementation.

We were able to run experiments with POPE up to 100 million entries, limited
only by the storage space available on our test machine.  We observed no
significant change in per-operation performance after one million entries,
indicating our construction should scale well to even larger datasets.

\begin{figure}[t]
\includegraphics[width=\linewidth]{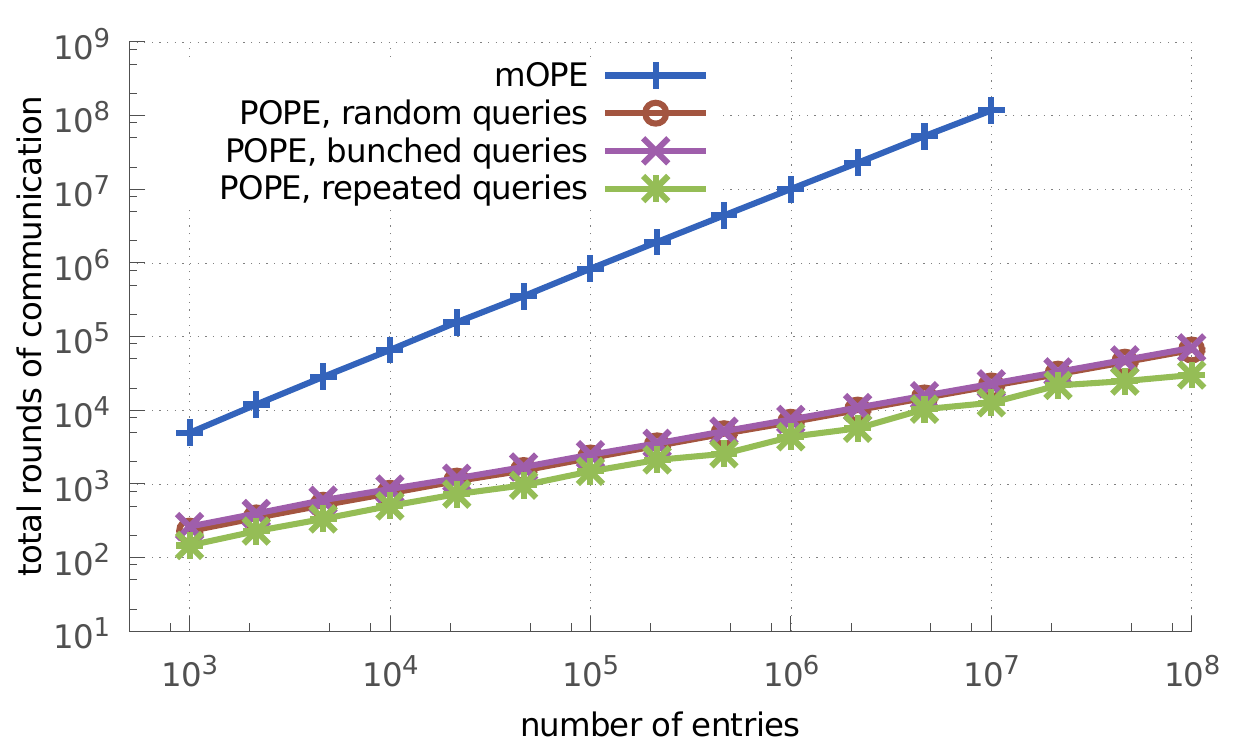}
\caption{Total rounds of communication for POPE and mOPE, plotted in log/log
scale according to total number of insertions $n$. 
Lower is better.
The number of range
queries in all cases was $\sqrt{n}$.\label{fig:rounds}}
\end{figure}

\begin{figure}[t]
\centering\includegraphics[width=\linewidth]{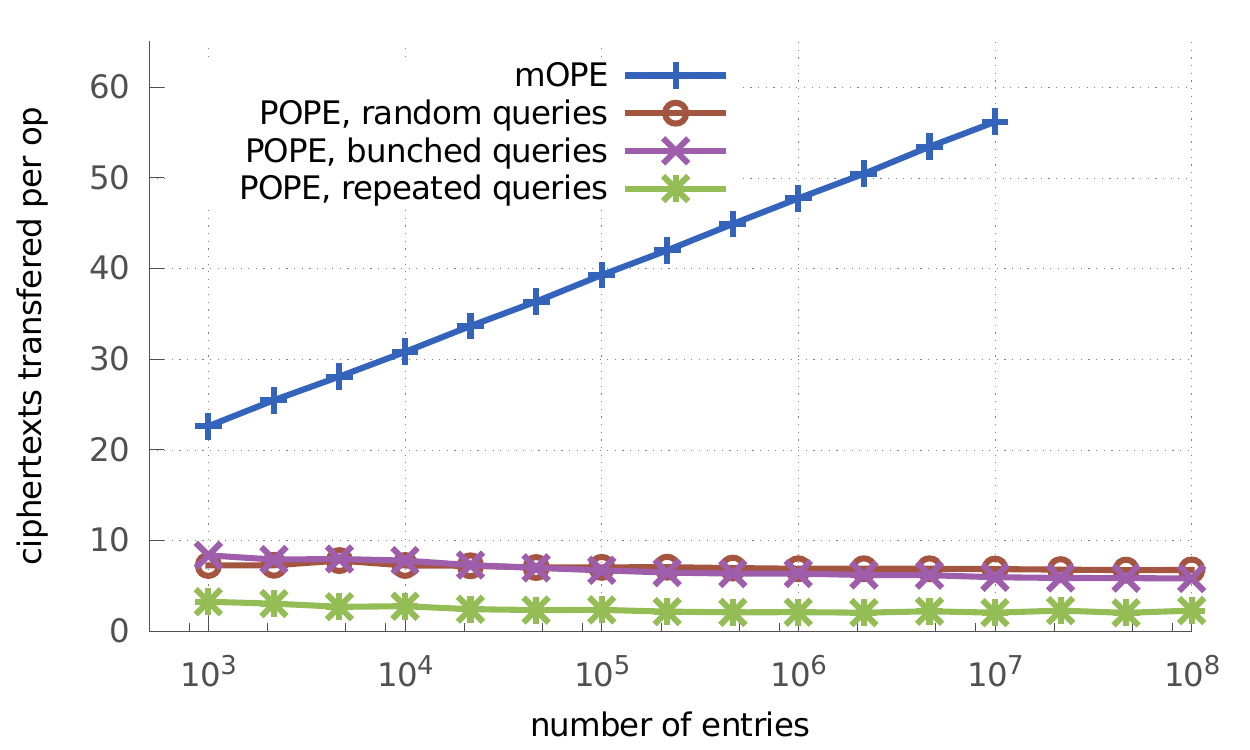}
\caption{Amortized communication costs for POPE and mOPE,
according to total number of insertions $n$. 
Lower is better.
The number of range
queries in all cases was $\sqrt{n}$.\label{fig:comm}}
\end{figure}

\subsection{Local Setting}

\paragraph{Experimental communication costs.}
Figures \ref{fig:rounds} and \ref{fig:comm} show the communication
costs, the total number of rounds of communication, and the average
number of ciphertexts transferred per operation. The number of
insertions $n$ is shown in the plots, and for each experiment we
performed $m=\sqrt{n}$ searches allowing $L=n^{1/4}$ entries stored in
temporary memory on the client.

As these figures demonstrate, the round complexity for POPE,
which is constant \emph{per range query}, is several orders of
magnitude less than that of mOPE. Furthermore, when averaged over all
operations, the number of ciphertexts transferred per operation for POPE
is roughly 7 in the worst case, whereas for mOPE this increases
logarithmically with the database size.

\begin{figure}[t]
\includegraphics[width=\linewidth]{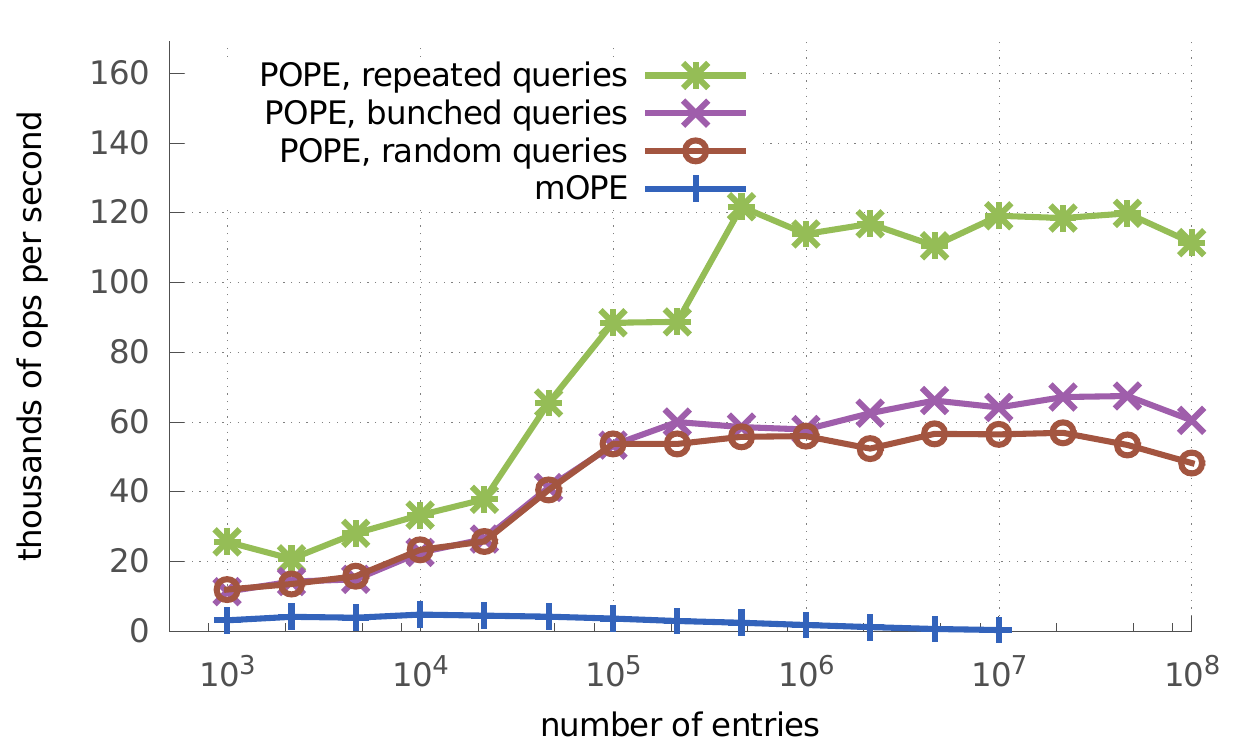}
\caption{Operations performed per second for POPE and mOPE.
Higher is better.
The number of range queries in all cases was $\sqrt{n}$.%
\label{fig:ops}}
\end{figure}

\begin{figure}
\centering\includegraphics[width=0.9\linewidth]{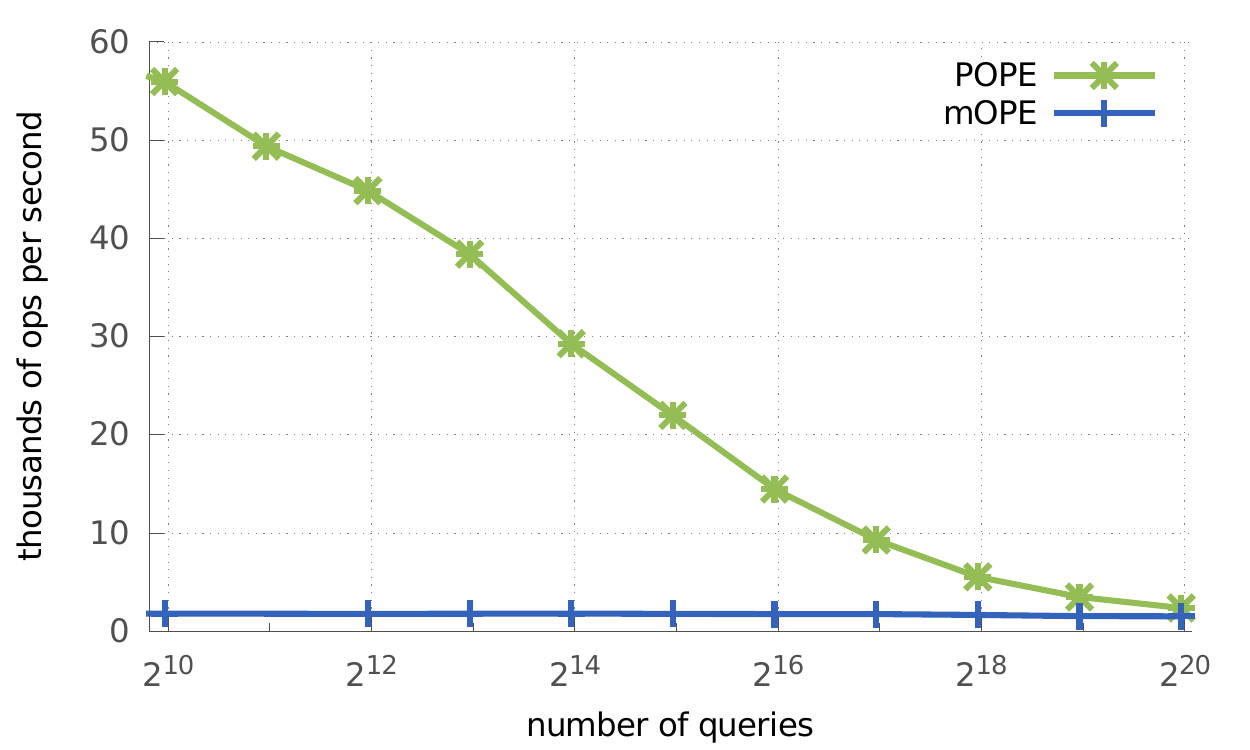}
\caption{Degradation in POPE performance with increasing number of
queries, measured in operations per second. Higher is better.
In all experiments, the number of insertions $n$ was fixed at 1
million, and the client-side storage at $L=32$. For these choices, $2^{10} \approx \sqrt{n}$ queries
is as shown in prior figures, and our $O(1)$-cost analysis holds up to $m \approx 2^{15}$.%
\label{fig:varym}}
\end{figure}

\paragraph{Experimental running time.}
The per-second operations counts, for our main experiments
with $n$ insertions, $m=\sqrt{n}$ range queries, and $L=n^{1/4}$
client-side storage, are presented in Figure~\ref{fig:ops}. 
For POPE, the performance increases until roughly 1 million entries,
after which the per-operation performance holds steadily between 50,000
operations per second with random, distinct queries, and 110,000
operations per second with a single, repeated query.

For one million entries and using our Python implementation without
parallelization, we achieved over 55,000 operations per second with POPE vs.
less than 2,000 operations per second for mOPE, without even accounting for the
network communication. 

Our POPE construction is well-suited particularly for problems
with many more insertions than range queries; indeed, the $O(1)$
theoretical performance guarantees hold only when $mL<n$.
Figure~\ref{fig:varym} shows the effects of varying numbers of range
queries on POPE performance. 
Although the performance of POPE clearly degrades with increasing
numbers of queries performed, this experiment shows competitive
performance compared to mOPE even when $m=n$.

\subsection{Experimental Network Effects}
We tested the effects of varying network latency and bandwidth using the
California public 
employees payroll data as described above. Our workload consisted of all
2,351,103 insertions as well as 1,000 random range queries at random
points throughout the insertions. Each range query result size was fixed
at 100 entries.

\begin{figure}[t]
\includegraphics[width=0.9\linewidth]{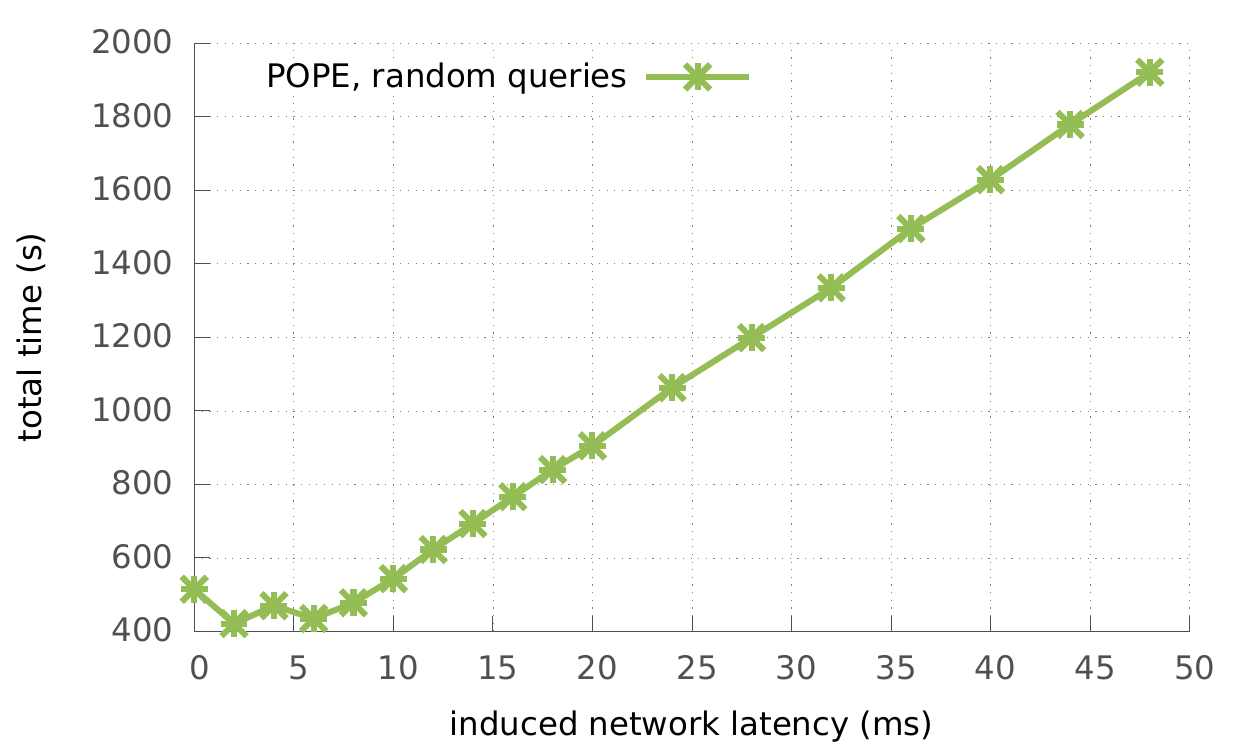}
\caption{Total running time for 2.3 million insertions and 1000 random
range queries, running over a network with varying artificially-induced
latency times.\label{fig:latency}}
\end{figure}

\begin{figure}[t]
\includegraphics[width=0.9\linewidth]{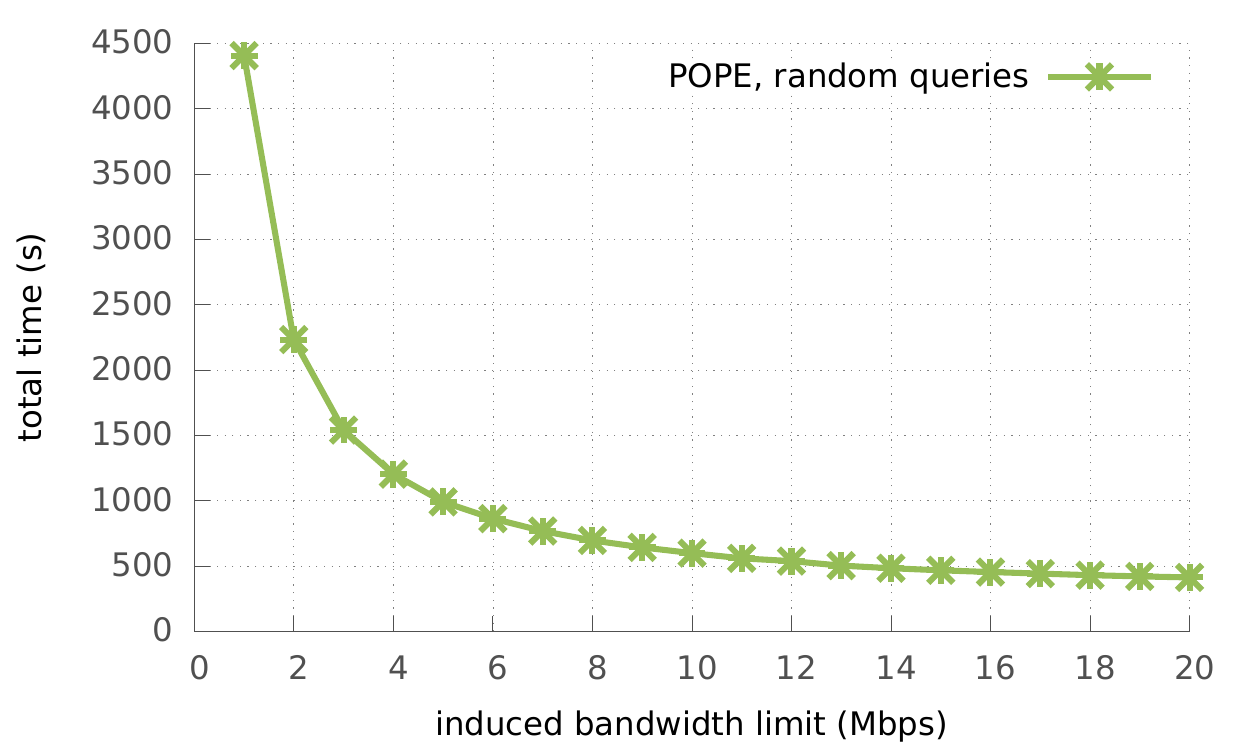}
\caption{Total running time for 2.3 million insertions and 1000 random
range queries, running over a network with varying bandwidth
limitations.\label{fig:bandwidth}}
\end{figure}

\Cref{fig:latency} shows the effects of latency on the POPE
implementation. With less than 5ms of latency, the cost is dominated by
that of the POPE computation and other overhead. Beyond this level, the
total runtime scales linearly with the latency. Note that 10ms to 30ms
represents typical server response times within the same continent over
the Internet.

\Cref{fig:bandwidth} shows the effects of bandwidth limitations on our
construction. Without any latency restriction, we limited the bandwidth
between 1 and 20 megabits per second (Mbps), which is the typical range
for 4G (on the low end) and home broadband Internet (on the high end)
connections. We can see that, past roughly 10 Mbps, the other overhead
of the implementation begins to dominate and there is no more
significant gain in speed.

\section{Related Work}

\paragraph{Order-Preserving and Order-Revealing Encryption.}
Order-preserving encryption (OPE)~\cite{AKSX04,BCLO09,BCO11} guarantees that $\enc(x) < \enc(y)$
iff $x < y$.  Thus, range queries can be performed directly over the
ciphertexts in the same way that such a query would be performed over
the plaintext data.  However, OPE comes with a security cost.  
None of the original schemes~\cite{AKSX04,BCLO09}
achieve the \emph{ideal} security goal for OPE of IND-OCPA (indistinguishability
under ordered chosen-plaintext attack)~\cite{BCLO09} in which ciphertexts
reveal no additional information beyond the order of the plaintexts.  In fact
Boldyreva et al.~\cite{BCLO09} prove that achieving a stateless encryption
scheme with this security goal is impossible under reasonable assumptions.  The
existing schemes, instead, either lack formal analysis or strive for weaker
notions of security which have been shown to reveal significant amount of
information about the plaintext~\cite{BCO11}.  The first scheme to achieve IND-OCPA 
security was the order-preserving encoding scheme of Popa et al.~\cite{PLZ13}.

A related primitive to OPE is order-revealing encryption (ORE) \cite{BCLO09},
which provides a public mechanism for comparing two encrypted values
and thus also enables range searches over encrypted data.
(Note, OPE is the special case where this mechanism is lexicographic
comparison,) 
The first construction of ORE satisfying \emph{ideal}
security~\cite{EC:BLRSZZ15} was based on multi-linear
maps~\cite{EC:GarGenHal13} and is thus unlikely to be practical in the
near future.  An alternative scheme based only on pseudorandom
functions~\cite{CLWW15}, however, has additional leakage that weakens
the achieved security.  

\paragraph{OPE alternatives.}
In addition to OPE there are several other lines of work that enable searching
over encrypted data.  Typically, these works provide stronger security
than provided by OPE; in particular they do not reveal the full order of
the underlying data as happens with OPE.  However, the additional
security guarantees come at a significant performance cost with even the latest schemes being one to two orders of magnitude slower than the latest OPE-based implementations~\cite{PKVKMCGKB14, ESORICS:FJKNRS15}.

Symmetric searchable encryption (SSE) was first proposed by Song, Wagner, and Perrig~\cite{SWP00} who showed how to search over encrypted data for keyword matches in sub-linear time.
The first formal security definition for SSE was given by Goh~\cite{Goh03}, Curtmola et
al.~\cite{CGKO06} showed the first SSE scheme with sublinear search time and compact space, while Cash et al.~\cite{CJJKRS13}
showed the first SSE scheme with sublinear search time for conjunctive queries.
Recent works~\cite{PKVKMCGKB14, CJJKRS13, ESORICS:FJKNRS15} achieve
performance within a couple orders of magnitude of unencrypted databases for
rich classes of queries including boolean formulas over keyword, and range
queries.  Of particular interest is the work of Hahn and Kerschbaum~\cite{CCS:HahKer14} who show how to use lazy techniques to build SSE with quick updates.  We refer interested readers to the survey by B{\"o}sch et al.~\cite{BHJP14} for an excellent overview of this area.  

Oblivious RAM~\cite{STOC:Goldreich87,AC:SCSL11,CCS:SDSFRY13,WCS15} and oblivious
storage schemes \cite{GMOT12,PKC:AKST14,DDF16,TTO15} can be used for the same
applications as OPE and POPE, but achieve a stronger security definition that
additionally hides the access pattern, and therefore incur a larger performance
cost than our approach.

Finally, we note that techniques such as fully-homomorphic
encryption~\cite{Gentry09}, public-key searchable
encryption~\cite{EC:BDOP04,BW07,SBCSP07}, and secure multi-party
computation~\cite{Yao86, BGW88, GMW86} can enable searching over encrypted data
while achieving the strongest possible security.  However, these approaches
    would require performing expensive cryptographic operations over the entire
    database on each query and are thus prohibitively expensive.  Very recently
    cryptographic primitives such as order-revealing
    encryption~\cite{EC:BLRSZZ15},
    as well as garbled random-access memory~\cite{STOC:GLOS15}, have offered the potential to achieve this level of security for sub-linear time search.  However, all constructions of these primitive either rely on very non-standard assumptions or are prohibitively slow.   

\paragraph{Lazy data structures and I/O complexity.}
Our POPE tree is is similar in concept to the Buffer Tree of \cite{Arg03}.
Their data structure delays insertions and searches in a buffer stored at each
node, which are cleared (thus executing the actual operations) when they become
sufficiently full. The main difference here is that our buffers contain only
insertions, and they are cleared only when a search operation passes through
that node.

We also point out an interesting connection to I/O complexity regarding
the size of local storage. In our construction, as in \cite{PRZB11}, the
client is treated as an oracle to perform comparisons of ciphertexts. If
we think of the client's memory as a ``working space'' of size $L$, and
the server's memory as external disk, then from
\cite{AV87} it can be seen that performing $m$ range queries on a
database of size $n\ge m$ requires a total transfer bandwidth of at
least $\Omega(m \log_L m)$ ciphertexts. (This is due to the lower bound
on the I/O complexity of sorting, and the fact that $m$ range queries
can reveal the order of a size-$m$ subset.)
In particular, this
means that the mOPE construction from \cite{PRZB11} cannot be improved
without either limiting the number of queries, or increasing the
client-side storage, both of which we do for POPE.

\paragraph{Acknowledgments}
We thank Jonathan Katz and David Cash for recommending the importance of the improved security of POPE.  We also thank the anonymous reviewers for their useful comments.

Daniel S. Roche's work is supported in part by Office of Naval Research
(ONR) award N0001416WX01489 and National Science Foundation (NSF) awards \#1319994 and \#1618269. 
Daniel Apon's work is supported in part by NSF awards \#1111599, \#1223623, and \#1514261. 
Seung Geol Choi's work is supported in part by ONR awards N0001416WX01489 and
N0001416WX01645, and NSF award \#1618269. 
Arkady Yerukhimovich's work is sponsored by the Assistant Secretary of Defense
for Research and Engineering under Air Force Contract No. FA8721-05-C-0002
and/or FA8702-15-D- 0001. Any opinions, findings, conclusions or
recommendations expressed in this material are those of the author(s) and do
not necessarily reflect the views of the Assistant Secretary of Defense for
Research and Engineering.

{\small
\bibliographystyle{plain}
\bibliography{POPE}

\begin{thebibliography}{10}

\bibitem{caldat14}
{California} public employee payroll data, 2014.
\newblock Source: Transparent California,
  \url{http://transparentcalifornia.com/downloads/}.

\bibitem{AV87}
Alok Aggarwal and Jeffrey~Scott Vitter.
\newblock The {I/O} complexity of sorting and related problems.
\newblock In {\em ICALP 1987}, volume 267 of {\em LNCS}, pages 467--478.
  Springer Berlin Heidelberg, 1987.

\bibitem{AKSX04}
Rakesh Agrawal, Jerry Kiernan, Ramakrishnan Srikant, and Yirong Xu.
\newblock Order-preserving encryption for numeric data.
\newblock In {\em ACM SIGMOD 2014}, pages 563--574, 2004.

\bibitem{PKC:AKST14}
Daniel Apon, Jonathan Katz, Elaine Shi, and Aishwarya Thiruvengadam.
\newblock Verifiable oblivious storage.
\newblock In {\em PKC~2014}, volume 8383 of {\em {LNCS}}, pages 131--148.
  Springer, Heidelberg, March 2014.

\bibitem{Arg03}
Lars Arge.
\newblock The buffer tree: A technique for designing batched external data
  structures.
\newblock {\em Algorithmica}, 37(1):1--24, 2003.

\bibitem{BGW88}
Michael Ben{-}Or, Shafi Goldwasser, and Avi Wigderson.
\newblock Completeness theorems for non-cryptographic fault-tolerant
  distributed computation (extended abstract).
\newblock In {\em 20th ACM STOC}, pages 1--10, 1988.

\bibitem{BCLO09}
Alexandra Boldyreva, Nathan Chenette, Younho Lee, and Adam O'Neill.
\newblock Order-preserving symmetric encryption.
\newblock In {\em EUROCRYPT~2009}, pages 224--241, 2009.

\bibitem{BCO11}
Alexandra Boldyreva, Nathan Chenette, and Adam O'Neill.
\newblock Order-preserving encryption revisited: Improved security analysis and
  alternative solutions.
\newblock In {\em CRYPTO~2011}, pages 578--595, 2011.

\bibitem{EC:BDOP04}
Dan Boneh, Giovanni {Di Crescenzo}, Rafail Ostrovsky, and Giuseppe Persiano.
\newblock Public key encryption with keyword search.
\newblock In {\em EUROCRYPT~2004}, volume 3027 of {\em {LNCS}}, pages 506--522.
  Springer, Heidelberg, May 2004.

\bibitem{EC:BLRSZZ15}
Dan Boneh, Kevin Lewi, Mariana Raykova, Amit Sahai, Mark Zhandry, and Joe
  Zimmerman.
\newblock Semantically secure order-revealing encryption: Multi-input
  functional encryption without obfuscation.
\newblock In {\em EUROCRYPT~2015, Part II}, volume 9057 of {\em {LNCS}}, pages
  563--594. Springer, Heidelberg, April 2015.

\bibitem{BW07}
Dan Boneh and Brent Waters.
\newblock Conjunctive, subset, and range queries on encrypted data.
\newblock In {\em TCC~2007}, pages 535--554, 2007.

\bibitem{BHJP14}
Christoph B{\"{o}}sch, Pieter~H. Hartel, Willem Jonker, and Andreas Peter.
\newblock A survey of provably secure searchable encryption.
\newblock {\em {ACM} Comput. Surv.}, 47(2):18:1--18:51, 2014.

\bibitem{CJJKRS13}
David Cash, Stanislaw Jarecki, Charanjit~S. Jutla, Hugo Krawczyk,
  Marcel{-}Catalin Rosu, and Michael Steiner.
\newblock Highly-scalable searchable symmetric encryption with support for
  boolean queries.
\newblock In {\em CRYPTO~2013, Part I}, pages 353--373, 2013.

\bibitem{CDGHWBCFG06}
Fay Chang, Jeffrey Dean, Sanjay Ghemawat, Wilson~C. Hsieh, Deborah~A. Wallach,
  Michael Burrows, Tushar Chandra, Andrew Fikes, and Robert Gruber.
\newblock Bigtable: {A} distributed storage system for structured data.
\newblock In {\em OSDI 2006}, pages 205--218, 2006.

\bibitem{CLWW15}
Nathan Chenette, Kevin Lewi, Stephen~A. Weis, and David~J. Wu.
\newblock Practical order-revealing encryption with limited leakage.
\newblock In {\em FSE}, pages 474--493. Springer, 2016.

\bibitem{CGKO06}
Reza Curtmola, Juan~A. Garay, Seny Kamara, and Rafail Ostrovsky.
\newblock Searchable symmetric encryption: improved definitions and efficient
  constructions.
\newblock In {\em ACM CCS 06}, pages 79--88, 2006.

\bibitem{DHJKLPSVV07}
Giuseppe DeCandia, Deniz Hastorun, Madan Jampani, Gunavardhan Kakulapati,
  Avinash Lakshman, Alex Pilchin, Swaminathan Sivasubramanian, Peter Vosshall,
  and Werner Vogels.
\newblock Dynamo: {Amazon}'s highly available key-value store.
\newblock In {\em SOSP 2007}, pages 205--220, 2007.

\bibitem{DDF16}
Srinivas Devadas, Marten van Dijk, Christopher~W. Fletcher, Ling Ren, Elaine
  Shi, and Daniel Wichs.
\newblock {Onion ORAM}: A constant bandwidth blowup oblivious {RAM}.
\newblock Theory of Cryptography Conference, {TCC} '16, 2016.

\bibitem{ESORICS:FJKNRS15}
Sky Faber, Stanislaw Jarecki, Hugo Krawczyk, Quan Nguyen, Marcel-Catalin Rosu,
  and Michael Steiner.
\newblock Rich queries on encrypted data: Beyond exact matches.
\newblock In {\em ESORICS~2015, Part II}, volume 9327 of {\em {LNCS}}, pages
  123--145. Springer, Heidelberg, September 2015.

\bibitem{EC:GarGenHal13}
Sanjam Garg, Craig Gentry, and Shai Halevi.
\newblock Candidate multilinear maps from ideal lattices.
\newblock In {\em EUROCRYPT~2013}, volume 7881 of {\em {LNCS}}, pages 1--17.
  Springer, Heidelberg, May 2013.

\bibitem{STOC:GLOS15}
Sanjam Garg, Steve Lu, Rafail Ostrovsky, and Alessandra Scafuro.
\newblock Garbled {RAM} from one-way functions.
\newblock In {\em 47th ACM STOC}, pages 449--458. {ACM} Press, June 2015.

\bibitem{Gentry09}
Craig Gentry.
\newblock Fully homomorphic encryption using ideal lattices.
\newblock In {\em 41st ACM STOC}, pages 169--178, 2009.

\bibitem{Goh03}
Eu{-}Jin Goh.
\newblock Secure indexes.
\newblock {\em {IACR} Cryptology ePrint Archive}, 2003:216, 2003.

\bibitem{STOC:Goldreich87}
Oded Goldreich.
\newblock Towards a theory of software protection and simulation by oblivious
  {RAMs}.
\newblock In {\em 19th ACM STOC}, pages 182--194. {ACM} Press, May 1987.

\bibitem{GMW86}
Oded Goldreich, Silvio Micali, and Avi Wigderson.
\newblock Proofs that yield nothing but their validity and a methodology of
  cryptographic protocol design (extended abstract).
\newblock In {\em 27th FOCS}, 1986.

\bibitem{GMOT12}
Michael~T. Goodrich, Michael Mitzenmacher, Olga Ohrimenko, and Roberto
  Tamassia.
\newblock Practical oblivious storage.
\newblock In {\em ACM CODASPY '12}, pages 13--24, 2012.

\bibitem{CCS:HahKer14}
Florian Hahn and Florian Kerschbaum.
\newblock Searchable encryption with secure and efficient updates.
\newblock In {\em ACM CCS 14}, pages 310--320, 2014.

\bibitem{quicksort}
C.~A.~R. Hoare.
\newblock Algorithm 64: Quicksort.
\newblock {\em Commun. ACM}, 4(7):321--, July 1961.

\bibitem{K15}
Florian Kerschbaum.
\newblock Frequency-hiding order-preserving encryption.
\newblock In {\em ACM CCS 15}, pages 656--667, 2015.

\bibitem{KS14}
Florian Kerschbaum and Axel Schr{\"{o}}pfer.
\newblock Optimal average-complexity ideal-security order-preserving
  encryption.
\newblock In {\em ACM CCS 14}, pages 275--286, 2014.

\bibitem{LLWB14}
Rui Li, Alex~X. Liu, Ann~L. Wang, and Bezawada Bruhadeshwar.
\newblock Fast range query processing with strong privacy protection for cloud
  computing.
\newblock {\em Proc. VLDB Endow.}, 7(14):1953--1964, October 2014.

\bibitem{TTO15}
Tarik Moataz, Travis Mayberry, and Erik-Oliver Blass.
\newblock Constant communication {ORAM} with small blocksize.
\newblock In {\em ACM CCS 15}, pages 862--873, 2015.

\bibitem{NKW15}
Muhammad Naveed, Seny Kamara, and Charles~V. Wright.
\newblock Inference attacks on property-preserving encrypted databases.
\newblock In {\em ACM CCS 15}, 2015.

\bibitem{PKVKMCGKB14}
Vasilis Pappas, Fernando Krell, Binh Vo, Vladimir Kolesnikov, Tal Malkin,
  Seung~Geol Choi, Wesley George, Angelos~D. Keromytis, and Steve Bellovin.
\newblock {Blind Seer}: {A} scalable private {DBMS}.
\newblock In {\em 2014 {IEEE} Symposium on Security and Privacy}, pages
  359--374, 2014.

\bibitem{PLZ13}
Raluca~A. Popa, Frank~H. Li, and Nickolai Zeldovich.
\newblock An ideal-security protocol for order-preserving encoding.
\newblock In {\em 2013 {IEEE} Symposium on Security and Privacy}, pages
  463--477, 2013.

\bibitem{PRZB11}
Raluca~A. Popa, Catherine M.~S. Redfield, Nickolai Zeldovich, and Hari
  Balakrishnan.
\newblock {CryptDB}: protecting confidentiality with encrypted query
  processing.
\newblock In {\em SOSP 2011}, pages 85--100, 2011.

\bibitem{SBCSP07}
Elaine Shi, John Bethencourt, Hubert~T.{-}H. Chan, Dawn~Xiaodong Song, and
  Adrian Perrig.
\newblock Multi-dimensional range query over encrypted data.
\newblock In {\em 2007 {IEEE} Symposium on Security and Privacy}, pages
  350--364, 2007.

\bibitem{AC:SCSL11}
Elaine Shi, T.-H.~Hubert Chan, Emil Stefanov, and Mingfei Li.
\newblock Oblivious {RAM} with $o((\log n)^3)$ worst-case cost.
\newblock In {\em ASIACRYPT~2011}, volume 7073 of {\em {LNCS}}, pages 197--214.
  Springer, Heidelberg, December 2011.

\bibitem{SWP00}
Dawn~Xiaodong Song, David Wagner, and Adrian Perrig.
\newblock Practical techniques for searches on encrypted data.
\newblock In {\em 2000 {IEEE} Symposium on Security and Privacy}, pages 44--55,
  2000.

\bibitem{CCS:SDSFRY13}
Emil Stefanov, Marten van Dijk, Elaine Shi, Christopher~W. Fletcher, Ling Ren,
  Xiangyao Yu, and Srinivas Devadas.
\newblock Path {ORAM}: an extremely simple oblivious {RAM} protocol.
\newblock In {\em ACM CCS 13}, pages 299--310. {ACM} Press, November 2013.

\bibitem{Accumulo}
{The Apache Software Foundation}.
\newblock Accumulo.
\newblock https://accumulo.apache.org/.
\newblock Accessed: 2015-09-24.

\bibitem{Cassandra}
{{The Apache Software Foundation}}.
\newblock Cassandra.
\newblock https://cassandra.apache.org/.
\newblock Accessed: 2015-09-24.

\bibitem{HBASE}
{{{The Apache Software Foundation}}}.
\newblock Hbase.
\newblock http://hbase.apache.org/.
\newblock Accessed: 2015-09-24.

\bibitem{WCS15}
Xiao Wang, Hubert Chan, and Elaine Shi.
\newblock Circuit {ORAM}: On tightness of the {Goldreich}-{Ostrovsky} lower
  bound.
\newblock In {\em ACM CCS 15}, pages 850--861, 2015.

\bibitem{WGA06}
D.~Westhoff, J.~Girao, and M.~Acharya.
\newblock Concealed data aggregation for reverse multicast traffic in sensor
  networks: Encryption, key distribution, and routing adaptation.
\newblock {\em Mobile Computing, IEEE Transactions on}, 5(10):1417--1431, Oct
  2006.

\bibitem{Yao86}
Andrew~Chi{-}Chih Yao.
\newblock How to generate and exchange secrets (extended abstract).
\newblock In {\em 27th FOCS}, pages 162--167, 1986.

\end{thebibliography}
}

\section{Proof of Theorem~1} \label{sec:app-analysis}
Choose $n, \epsilon$ so that $L = n^\epsilon > 16$.
The case of $\mathsf{Insert}$ is trivial to analyze: The server never makes
comparison requests.  So, we focus on the case of $\mathsf{Search}$.

\medskip
\noindent{\bf An alternative $\mathsf{split}$ procedure}
To simplify our analysis, we introduce an alternative version of the leaf splitting procedure which discards any split that results in very unbalanced partitions.  We argue that such a split will always (in expectation) be worse than our original split and thus can be used to bound its costs.

Let $z=2L/\log{L}$.
We say that a set of $L$ pivot points is \emph{$z$-balanced} if there are $z$ (out of $L$) pivots such that partitioning a node of size $k$ using these $z$ pivots only results in partitions that are each of size at most $2k/z$.

\begin{framed}
\small
\begin{enumerate}
\item[A.] 
As in actual $\mathsf{Split}$, choose $L$ pivots uniformly at random and partition the labels according to these pivots.
\item[B.] If the $L$ pivots are not $z$-balanced, throw out the partition and recurse on the same node.
\item[C.]
If these $L$ pivots are $z$-balanced, promote only the $z$ balanced pivots (instead of the total $L$) to the parent, partition the labels and recurse on the node containing the searched label.
\end{enumerate}
\end{framed}
We argue that this procedure is worse than the original split both in
the (expected) number of rounds and the (expected) total bandwidth (over
all $m$ queries).  To see this for the number of rounds, observe that
the alternate procedure chooses its partitions in the same way as the
original, but always drops some (or all) of the pivot points resulting
in larger nodes and a deeper recursion to reach nodes of size $L$.  For
the case of bandwidth most of the cost comes from streaming labels to
the client to partition them when splitting a leaf, which takes $O(k)$
bandwidth for a node of size $k$.  Now consider a single label $x$ in
the tree.  This may get moved between leaf nodes multiple times during the queries, but
each time it is moved the node it lands in is larger if the alternative
split procedure is used as argued above, as compared to the actual
\Psplit.  Thus, the total cost of all
splits over $m$ queries is larger for the alternative split as it will
require repeatedly streaming these larger nodes to the client.  

For the remainder of this proof, we analyze the alternative procedure for
splitting a leaf to bound the costs of the real one.  

\medskip
\noindent {\bf Round complexity for a single search.}
The round complexity for a search can be computed by considering the round
complexity for splitting at internal nodes (case (i) in
Section~\ref{sec:protocol}) and splitting at a leaf (case (ii) in
Section~\ref{sec:protocol}). 

The round complexity for case (i) is asymptotically the same as the height of the
tree. Since the tree is re-balanced such that each internal node contains at
least $L/2$ labels, the height of the tree is $O(\log_L n)$. 

{ 
As for case (ii), we first need to show that $L$ random pivots are $z$-balanced with constant probability, so that there is a successful split after $O(1)$ many unsuccessful ones.  
To see this, define an imaginary sorted
list $(X_1, \ldots, X_{z})$ that contains the $k$ input labels in sorted
order, equally partitioned so each $X_i$ has $k/z$ elements. 
Note if each $X_i$ contains at least one pivot (out of the chosen $L$
pivots), then the pivots must be $z$-balanced; in particular, one can
\unskip\parfillskip 0pt \par} 

\vfill\eject

\noindent
find such pivots by choosing
one from each $X_i$.
By the Coupon Collector's Problem, the probability that $L$ pivots hit all the
$X_i$s is constant.  

Now, note that, by the definition of $z$-balanced, after each successful split the size of the largest partition is reduced by 
a factor of $z/2=L/\log{L}$.  Thus, the total number of successful
splits needed and also the total (expected) recursion depth is
$O(\log_{L/\log L} n)$, which simplifies to $O(\log_L n)$ when $n\ge 16$.
The total round complexity of the POPE protocol is therefore
$O(\log_L{n})$.  

\medskip
\noindent {\bf Total bandwidth over $m$ search queries.}

\noindent {\it Height of the tree.}
First, we need a tighter analysis on the height of the POPE tree. For this, we start with
counting the total number of labels in the internal nodes. 
The total number of $\mathsf{Split}$ calls over all $m$ $\mathsf{Search}$
operations is at most $O(m\log_L n)$, since each search has $O(\log_L n)$
recursion depth. 

Now, consider the sorted labels in non-leaf nodes of the tree.
Each such label is inserted by a $\mathsf{Split}$ operation from a leaf, and
each $\mathsf{Split}$ inserts at most $z$ labels. Therefore, the total number of labels
stored in the sorted, non-leaf portion of the tree $\mathcal{T}$ is 
$O(z m\log_L n)$, which is $O(mL\log n)$. 
Recall the sorted labels in the non-leaf nodes of the tree form a $B$-tree with
between $L$ and $L/2$ labels per node (after rebalancing).  Therefore, the
maximum height of the tree is 
$\mathsf{height}(\mathcal{T}) = O\big(\log_L (mL\log n) \big )$.

\medskip
\noindent {\it Sending sorted labels to client.}
Recall that the round complexity of a search is
$O(\log_L n)$. Each round of $\mathsf{Search}$ involves uploading at most
$L$ labels to serve as partition indices to the client, incurring a total bandwidth of
$B_{\ell} = O(mL\log_L{n})$. 

\medskip
\noindent {\it Sending labels in non-leaf buffers to client.}
In addition, all the labels in buffers along the search path are sent to the
client -- some more than once. Observe that labels in buffers only move to a
lower buffer, or laterally from leaf nodes to leaf nodes during $\mathsf{Split}$
operations, which means that any label in 
{\em non-leaf} nodes must be sent to the client at most $\mathsf{height}(\TTT)$
times. 
Therefore, the expected total bandwidth for the labels in \emph{non-leaf}
buffers, across all $\mathsf{Search}$ operations, is
$B_{in} = O(n\cdot\mathsf{height}(\mathcal{T})) = O(n\log_L{m} +
n\log_L(\log{n}))$.

\medskip
\noindent{\it Communication cost of splits.}
Observe that splitting a leaf node of size $k$, through all the recursive calls, takes bandwidth $O(k)$ since 
$O(k+k/z+k/z^2+\ldots)=O(k)$.  So, we consider the
total size of all leaf nodes encountered during $\mathsf{Search}$ operations to
compute the costs from splits. 
The worst-case scenario for the construction is when all $n$
insertions happen before all $m$ searches, and each search's splits land in the
largest remaining leaf node(s). Using the alternative split
procedure, the largest possible leaf
nodes the search's splits will land in (counting only successful splits as rounds) have the following sizes:
\begin{enumerate}
\small
\item[-] (Round 1) 1 node of size $n$. A split lands on this node, splitting it
    into nodes of size at most $n \cdot (2/z)$.
\item[-] (Round 2) $z$ nodes of size at most $n \cdot (2/z)$. Note the total size of the
    nodes is at most $n$. 
\item[-] (Round 3) $z^2$ nodes of size at most $n \cdot (2/z)^2$. The total size
    is at most $n$, etc. 
\end{enumerate}
We have $\sum_{i=0}^{w} z^i  \ge m$ with $w = O(\log_L m)$, and $m$ largest
leaf nodes are encountered by round $w$. Since the total size of the touched nodes in each
round is at most $n$, the total size of the $m$ largest leaf nodes is bounded by
$B_{s} = nw = O(n\log_L m)$. 

By summing up $B_\ell, B_{in}, B_s$, we find that the total bandwidth over all
$(n+m)$ operations is at most $O(mL\log_L{n} + n\log_L{m} + \allowbreak n\log_L(\log{n}))$, and
Theorem~\ref{thm:cost} follows.

\end{document}